# Ethnic conflicts, civil war and economic growth: Region-level evidence from former Yugoslavia

Aleksandar Kešeljević[†]     Stefan Nikolić[‡]     Rok Spruk[§]

May 5, 2025

**Abstract**

We investigate the long-term impact of civil war on subnational economic growth across 78 regions in five former Yugoslav republics from 1950 to 2015. Leveraging the outbreak of ethnic tensions and the onset of conflict, we construct counterfactual growth trajectories using a robust region-level donor pool from 28 conflict-free countries. Applying a hybrid synthetic control and difference-in-differences approach, we find that the war in former Yugoslavia inflicted unprecedented regional per capita GDP losses estimated at 38 percent, with substantial regional heterogeneity. The most war-affected regions suffered prolonged and permanent economic declines, while capital cities experienced more transitory effects. Our results are robust to extensive variety of specification tests, placebo analyses, and falsification exercises. Notably, ethnic tensions between Serbs and Croats explain up to 40 percent of the observed variation in economic losses, underscoring the deep and lasting influence of ethnic divisions on economic impacts of the armed conflicts.

**Keywords:** Ethnic conflict, civil war, economic growth, Yugoslavia
**JEL codes:** O1, N1, R1

[†]Associate Professor of Economics, School of Economics and Business, University of Ljubljana, Kardeljeva ploščad 17, SI-1000 Ljubljana. saso.keseljevic@ef.uni-lj.si.
[‡]Lecturer in Economics, Loughborough University, Epinal Way, LE11 3TU Loughborough, United Kingdom, s.nikolic@lboro.ac.uk.
[§]Associate Professor of Economics, School of Economics and Business, University of Ljubljana, Kardeljeva ploščad 17, SI-1000 Ljubljana, rok.spruk@ef.uni-lj.si.

We thank the participants of the 20th annual conference of the Italian Society of Law and Economics (Sapienza) for their invaluable feedback. Standard disclaimer applies.

# 1 Introduction

Wars and armed conflicts are widely recognized as principal impediments to sustainable development, inflicting profound human suffering and imposing substantial economic, social, and humanitarian costs (Keynes, 2017; Reynal-Querol, 2002; Collier, 2003; Bilmes and Stiglitz, 2008; Blattman and Miguel, 2010; Sachs, 2012; Hoeffler, 2017). While international and civil wars alike devastate civilian populations, civil conflicts often promulgate more severe and irreversible consequences due to widespread displacement, high civilian casualties, refugee crises, and long-term fiscal and social disruptions. These indirect costs, frequently difficult to quantify, compound the immediate destruction, leaving enduring wounds on affected societies.

Three methodological approaches dominate the empirical efforts to estimate economic costs of conflict. First, the cost accounting approach categorizes economic and social damages through relative price frameworks, capturing diverse direct and indirect impacts (Arunatilake et al., 2001; Bilmes and Stiglitz, 2006; Skaperdas, 2011). Second, time-series and panel regression analyses employ difference-in-differences estimators to isolate conflict effects from observable economic data (Enders et al., 1992; Barro and Lee, 1994). And third, more recently, the synthetic control method, pioneered by Abadie and Gardeazabal (2003) and refined by subsequent scholars (Abadie et al., 2015; Gilchrist et al., 2023), has emerged as a robust tool for constructing counterfactual growth and development scenarios of armed conflicts. By comparing conflict-affected regions with their synthetic counterparts, as a weighted combination of unaffected areas—this particular approach circumvents arbitrary model assumptions and extrapolation risks (Abadie, 2021). This approach has been applied to conflicts such as the Donbass war (Bluszcz and Valente, 2022), the economic consequences of Arab Spring and civil war in Libya (Echevarría and García-Enríquez, 2019), and the civil wars in former Yugoslavia (Kešeljević and Spruk, 2023) and Syria (Kešeljević and Spruk, 2024), among others.

This study examines the impact of the Yugoslav Civil War (1987–1995) on the long-term regional growth trajectories of former Yugoslav economies, leveraging the rise of military violence by the Yugoslav National Army in the 1980s as a source of variation. We provide granular, region-level evidence on how armed conflict shaped economic development from 1950 to 2015 in 78 Yugoslav regions. We apply the synthetic control method (Abadie and Gardeazabal, 2003) to construct a plausible counterfactual scenario to estimate the trajectory of economic growth that would have prevailed in the absence of conflict. We draw on a stable donor pool comprising regions from 28 countries over the same period that have not been affected by armed conflict. We find that the war in former Yugoslavia inflicted unprecedented regional per capita GDP losses estimated at 38 percent, with substantial regional heterogeneity. The most war-affected regions suffered prolonged and permanent economic declines, while capital cities experienced more transitory effects. The regional dimension of our data allows investigating the transmission mechanisms behind these economic effects. We find ethnic polarisation and population displacement to be the main correlates of regional per-capita GDP losses. To the best of our knowledge, this is the first study to explore subnational disparities in the economic impact of armed conflict using a multi-country, multi-regional donor pool and counterfactual methods.

Empirical studies conducted at the national or cross-national level are inherently limited in their ability to capture the differential effects of conflict across subnational units. Such analyses typically assume a homogeneity of impact across all regions, failing to account for the heterogeneous economic legacies that result from varying levels of exposure to conflict violence. This gap in the literature is particularly significant in the context of civil wars, as the economic consequences of conflict are often geographically concentrated and disproportionally affect specific regions. As Gates et al. (2012) argues, civil wars do not unfold uniformly across an entire country; rather, they are typically localized to particular subnational units, where the conflict is concentrated. These regions bear the brunt of the violence and the associated economic and social disruptions. Consequently, national or cross-national studies, which aggregate data across entire countries, obscure the important variation that exists at the subnational level, thus masking crucial insights into the long-term effects of civil war.

The subnational level offers a more nuanced perspective, allowing us to investigate the differential impact of the war on regions with varying levels of exposure to violence.

One key question that arises is whether all subnational units experience faster growth following the end of conflict, or if the recovery is uneven. As Brinkerhoff (2011) highlights, local governance structures are often better positioned to organize and manage post-conflict recovery efforts, which can lead to divergent outcomes across districts. Therefore, analyzing the subnational level provides a richer understanding of the specific factors that contribute to the varying pace and nature of recovery – factors that would be overlooked in a national-level analysis.

Moreover, it is crucial to examine whether districts that experienced more intense violence during the civil war exhibit slower economic recovery than those that were less affected. Despite the overall peace dividend benefiting all districts, it is possible that regions with higher levels of conflict exposure face long-lasting disadvantages in terms of economic growth. A subnational approach enables us to explore this question and assess whether there exists a relative growth disadvantage for areas that were most severely impacted by the conflict.

Despite extensive scholarship on the Yugoslav conflicts of the late 20th century, empirical assessments of their direct and indirect costs, particularly at regional levels or over the long term, remain scarce. The decade-long war in former Yugoslavia (1991–2001) resulted in approximately 130,000–140,000 civilian casualties, forcibly displaced 4 million individuals[1], and left around 33% of the population with severe post-traumatic stress disorder (PTSD) (Başoğlu et al., 2005). Notably, 64% of deaths were civilian (Khorram-Manesh et al., 2021) – a proportion considerably exceeding those of the Spanish Civil War (50%), World War II (48%), and Vietnam War (40%) – while perinatal mortality rates reached unprecedented heights (Fatusic et al., 2005).

By utilizing a large-scale subnational dataset of per capita GDP and its determinants, we compare the economic growth trajectories of Yugoslav regions before and after the outbreak of the civil war. Through a rigorous application of the hybrid synthetic control and difference-in-differences estimator, we estimate the counterfactual scenario of long-term economic growth by quantitatively reproducing the pre-war trajectories from the implicit attributes on unaffected regions outside former Yugoslavia, and comparing them to their actual counterparts.

---

[1] See: Report on Transitional Justice in the Former Yugoslavia. New York, NY: International Center for Transitional Justice (https://www.ictj.org/publication/transitional-justice-former-yugoslavia).

We perform a battery of robustness tests, including large-sample permutation inference, interactive fixed-effects algorithm, and in-time placebo analysis. By assigning the shock of war to unaffected regions outside Yugoslavia and comparing the placebo gaps with the estimated gaps, our approach pinpoints the uniqueness of the effect and detects whether the economic cost of civil war is uniquely perceivable in former Yugoslavia or not. Conversely, by assigning the shock to a deliberately false date, our approach credibly detects whether the estimated gaps are driven by pre-war economic difficulties, institutional sclerosis of former Yugoslavia and the lack or macroeconomic stabilization in the 1980s, or by other alternative shocks. Additionally, by employing more advanced treatment effect estimators, we further gauge and assess the robustness of our estimates against the relaxation of traditional assumptions in counterfactual analysis.

Our findings contribute to the extensive literature in economics on conflict, emphasizing the intensity, persistence and regional variation in the economic effect of the Yugoslav civil war, which appears to be large and long-lasting. Regardless of which estimates and models are used to assess the impact of armed conflicts, the effect of civil war on output and economic activity is always disastrous (Collier, 1999; Gates et al., 2012; Müller, 2013). We estimate an effect that is substantially larger than those found in the literature so far. On average, the Yugoslav conflict reduced regional per capita GDP by 38 percent, with the most affected regions losing as much as 79 percent in per capita GDP relative to their control groups. Compared to previous studies on the economic costs of armed conflict, we show that the intensity of the negative effect of war in former Yugoslavia is approximately one third higher compared to the average civil war, producing a long-lasting derailment of economic growth trajectories instead of the temporary effects most commonly detected in the literature on armed conflict. Furthermore, we find evidence of widespread and distinctive heterogeneity in the effects, which varies with ethnic and economic geography, levels of economic development, and historical legacy.

In the context of the Yugoslav conflict, we find that both forced migration and population displacement, as well as ethnic fractionalization and polarisation, were powerful determinants of regional GDP loss. These findings resonate with well-established, cross-country studies by Montalvo and Reynal-Querol (2005a,b), which find that ethnic polarization has a statistically significant effect on the incidence of civil wars, while ethnic fractionalization negatively affects economic development. Our results also complement recent, microeconometric studies that identify persistent effects of displacement on economic outcomes (Huber et al., 2021; Testa, 2021). Our study confirms the importance of these mechanisms, which we observe over the long term at the regional level. In this regard, our results emphasize that population displacement and forced migration are relatively more important drivers of regional economic losses from armed insurgency than ethnic fractionalization alone (Montalvo and Reynal-Querol, 2005a; Kondylis, 2010).

The rest of the paper is organized as follows. Section 2 reviews the relevant literature. Section 3 presents the contextual background behind the Yugoslav conflict. Section 4 presents the empirical strategy. Section 5 discusses data and samples. Section 6 discusses the results, including robustness checks and mechanisms. Section 7 offers our conclusions.

## 2  Literature Review

The economic consequences of civil war are universally devastating, often triggering what Collier (1999) has dubbed "development in reverse". To date, existing empirical studies consistently document the long-term economic stagnation and structural breakdown of growth trajectories in response to the armed conflict. The notion that civil wars inflict significant economic losses through multiple channels, including the destruction of key infrastructure, capital flight, displacement of human capital, and erosion of institutional fabric, can be only scarcely disputed. By way of example, five years of civil war lead to a decline in per capita GDP of between 7 and 22 percent (Müller, 2013). Beyond the immediate economic devastation, civil wars generate profound humanitarian crises, as evidenced by Müller's striking statistic that, on average, 22 individuals become refugees for every person killed in conflict. The impact is further heightened in ethnically polarized societies, where deep-seated divisions prolong instability and hinder post-conflict recovery.

Several recent studies have investigated the economic impact of armed conflict utilizing synthetic control methods. Horiuchi and Mayerson (2015) examine the economic consequences of the Second Palestinian Intifada (2000) on Israel's economy, applying the synthetic control method to project the counterfactual GDP trajectory. Their findings reveal a stark structural breakdown in Israel's growth path, with the synthetic version of Israel maintaining a permanently higher level of per capita GDP compared to its real counterpart. The estimated economic loss attributable to the Second Intifada amounts to approximately 9 percent of per capita GDP, underscoring the severe and lasting economic repercussions of prolonged conflict.

Echevarría and García-Enríquez (2020) investigate the economic costs of the Arab Spring, focusing on the Egyptian revolution of 2011. Their analysis identifies substantial cumulative losses in both real GDP per capita and its growth rate. The structural disruption induced by the revolution resulted in an estimated 1.5 percentage point reduction in Egypt's per capita GDP growth rate compared to the counterfactual scenario. The counterfactual estimation of economic growth trajectories in the hypothetical absence of armed conflict has become a central focus of contemporary research. Other recent studies have extended this framework to diverse conflict settings, including hybrid warfare and Russian separatism in Donbas (Bluszcz and Valente, 2022), the civil war in Syria (Kešeljević and Spruk, 2023), the rise of Hezbollah and sectarian clashes in Lebanon (Emery and Spruk, 2024), the overthrow of the Gaddafi regime and subsequent Libyan civil war (Echevarría and García-Enríquez, 2019), and the economic consequences of Kurdish separatism in Türkiye (Bilgel and Karahasan, 2019).

## 2.1 Mechanisms

The existing literature on the economic and social consequences of armed conflict has discovered a multitude of mechanisms through which war may impact economic growth.[2]

---

[2] For a recent literature review of the impacts of armed conflict on human development see Vesco et al. (2025).

War can destroy or displace physical and human capital. It may lower consumption, disincentivize investment, compress government spending, or limit trade. Consequently, shocks of war can influence both aggregate supply and demand. While most studies tend to identify a single mechanism within a specific conflict, some papers manage to uncover multiple mechanisms at play. In this subsection, we summarize the literature on some of the key mechanisms, including those that we can test in the Yugoslav context.

**Health**. Exposure to armed conflict is detrimental to physical and mental health. For instance, Akresh et al. (2012) find that people exposed to the Nigerian civil war as children and adolescents, exhibit reduced stature whilst Akresh et al. (2012) discover that children exposed to the Eritrea–Ethiopia war have lower height-for-age Z-scores, with similar effects for children born before or during the war. By analyzing food shortages during World War I in Germany, Blum (2013) shows that children from the middle and lower social class experienced the sharpest decline in average height. Gazeley and Newell (2013) also find evidence of stark decrease in the caloric value of foods and a reduction in the available quantities of crucial vitamins among workers in Britain during World War I. Furthermore, Chamarbagwala and Moránb (2011) estimate that Guatemala's civil war had a negative impact on human capital accumulation. In particular, the authors find a strong negative impact of the civil war on the education of rural communities, effectively deepening gender, regional, sectoral, and ethnic disparities in schooling. In terms of further example, Conzoab and Salustric (2019) find that individuals exposed to combat in the first six years of life display lower trust and social engagement well into adulthood. In addition, Kesternich et al. (2014) present compelling evidence, that those experiencing war or combat have significantly lower self-rated health as adults.

**Population displacement and forced migration.** Another powerful channel of war are displacements and forced migration movements. The consensus in the literature is that expulsions tend to generate negative outcomes both at the individual but also the aggregate level in the sending economies (Becker and Ferrara, 2019). Acemoglu et al. (2011) study the effect of the Holocaust on economic growth outcomes in Russia, showing that increased intensity in the persecution and mass murder of Jews by the Nazis reduced cities' population size and worsened economic and political outcomes since the collapse of the Soviet Union. Forced migration not only affects receiving and sending populations, but also migrants themselves. For example, Bauer et al. (2013) find that first-generation migrants in Germany who experienced displacement after World War II, still tend to be economically disadvantaged. Displaced agricultural workers, however, exhibit higher incomes than comparable natives, as displacement caused large-scale transitions out of low-paid agriculture. Similarly, Kondylis (2010) shows that displaced Bosnian men experience higher unemployment levels, and displaced women are more likely to drop out of the labor force.

**Mobilization**. Military conscription had an effect on a range of socio-economic outcomes. Jaworski (2014) shows compelling evidence that manpower mobilization during World War II decreased educational attainment among high school-age females during the early 1940s, even reduced employment and earnings, and altered decisions regarding family formation. Kholodilin (2016) shows that the outbreak of World War I led to an outflow of men from cities in Germany. Toward the end of the war, the construction freeze accompanied by an inflow of workers and discharged soldiers resulted in a housing shortage. Moreover, Castañeda Dower and Markevich (2018) find that mass mobilization of military draftees in Russia during World War I, led to a dramatic decrease of agricultural production, which was particularly pronounced among private farms. Based on the examination of perinatal mortality in Britain after the onset of World War II, Geiger and Wichert (2019) find that perinatal mortality increased immediately, which can be attributed to the decline in quality of medical care due to the sudden conscription of trained physicians. Cesur and Sabia (2016) find that combat assignments are associated with a substantial decline in relationship health and an increased risk of domestic violence, while Cousley et al. (2017) find some evidence of a small, temporary negative effect of military service on employment and a substantial positive effect on post-school qualifications below university level, home ownership and marriage.

**Gender imbalance.** War-induced gender imbalance has significant consequences on family outcomes, as well as the labor and marriage market. By way of example, Bethmann and Kvasnicka (2013) show that shortfalls of men after World War II significantly increased the non-marital fertility ratio. However, this effect is significantly lower in German counties with a larger share of prisoners of war. In line with this, Kesternich et al. (2020) find that the stark sex ratio imbalance induced by World War II led to a short-term pattern of decreased fertility rates in Germany. The authors find that female cohorts with low sex ratios have fewer children at younger ages and a larger fraction remains childless. Brainerd (2017) discovers that the post-World War II sex ratio imbalance in Russia led to persistent lower rates of marriage and fertility, higher nonmarital births, and reduced bargaining power within marriage for women most affected by war deaths.

**Trade**. Trade disruptions are an important channel through which war impacts the economy. Many studies discuss the disastrous effect of the Naval Blockade of Germany during World War I (Ritschl, 2005; Davis and Engerman, 2012; Böse and Ziegler, 2015). In particular, Osborne (2004) and Kramer (2014) emphasize that the blockade was the single most effective weapon at the disposal of the Allies during World War I, especially due to its impact on food and raw material supply. In this regard, Vincent (1985) offers a compelling description of the hunger, physical deformity, and death resulting from an extended period of malnutrition, and its impact on psychological suffering. Cox (2015) finds that as a consequence of the Hunger blockade between 1914 and 1924, German children, especially those from the working class, suffered severe malnutrition.

**Finance**. Some studies also investigate the reaction of financial markets and firms to war or conflict. Verdickt (2020) finds that both managers and investors became more risk averse as a consequence of news coverage on conflict before the outbreak of World War I. Schneider and Troeger (2006) show that the conflicts in 1990-2000, such as those in Iraq or Yugoslavia, affected the interactions at the core financial markets in the Western world negatively. Frey and Kucher (2000) find that historically significant events related to World War II are reflected in asset prices, especially government bond prices. Furthermore, Baten and Schulz (2005) provide evidence that most companies in Germany during World War I did not make high profits during the war, the only exception being the metal/machinery and chemical industries. The median entrepreneur however, experienced an income decline of a magnitude similar to the decline in the real wage of workers. Lastly, Liu (2020) finds that counties with greater pre-war exposure to international trade experienced more firm entry after World War I. The effect was delayed because the war simultaneously hindered machinery imports. Better access to finance also contributed to firm entry after the trade shock. Other authors analyse various aspects of war spending and war-related industry (Lafond et al., 2022). Fishback and Cullen (2013) find that in the longer-term counties receiving more war spending per capita during World War II experienced greater population growth. On the other hand, Jaworski (2017) argues that despite a boom in manufacturing activity during the war, the evidence is not consistent with differential postwar growth in counties that received more investment.

**Additional channels.** Some studies emphasize multiple channels through which armed conflict can impact socio-economic outcomes. Examining the economic effects of Donbass war, Bluszcz and Valente (2022) demonstrate that the war-induced disruption to production, trade and employment, agricultural and financial losses, compression of public expenditures, and a partial military mobilization coupled with growing political instability led to a loss in Ukraine's per capita GDP amounting to 15.1% on average for 2013–2017. Somewhat relatedly, Hönig (2021) shows that the civil war in Sierra Leone led to a higher share of workers in agriculture, fewer educated workers, and lower worker income. These effects can be attributed to direct effects on income and indirect effects by changing the allocation of labor across sectors and locations, changes in amenities, firm productivity losses and selective migration. Furthermore, Markevich and Harrison (2011) find that the Great War and Civil War produced the deepest economic trauma of Russia's troubled twentieth century, which can be attributed to different factors such as famine, wartime disruptions and confiscatory disincentives.

# 3   Contextual Background

The dissolution of Yugoslavia precipitated one of the most violent conflicts in modern European history, culminating in the first large-scale armed confrontation on the continent since World War II. A central explanation for the collapse of the Yugoslav federation and the subsequent civil war (late 1980s–1995) lies in deep-seated ethnic divisions and polarization, which had long shaped the political and social fabric of the region (Kaufman, 1996; Bianchini and Schöpflin, 1998; Petak, 2003; Ivekovic, 2000; Elzarka, 2018). While Slovenia remained relatively homogeneous, the rest of the Yugoslav republics exhibited significant ethnic heterogeneity. Ethnic diversity has been linked to heightened political instability and economic underperformance in autocratic regimes (Collier, 2000, 2001; Montalvo and Reynal-Querol, 2005a,b; Alesina and La Ferrara, 2005; Miladinović, 2024) opposed to democratic regimes, where political liberalization may effectively eliminate the adverse economic effects of ethnic diversity (Montalvo and Reynal-Querol, 2021).

The historical roots of ethnic tensions between Serbs and Croats trace back to World War II, when Nazi Germany invaded Yugoslavia. During this period, Croatia established a Nazi-aligned puppet state under the Ustasha regime, which engaged in systematic atrocities against Serbs, Jews, Roma, and other non-Croat populations, most notably in the Jasenovac concentration camp (Levy, 2009). In contrast, Serbia's Royal Army (Chetniks) aligned with the Allied forces, receiving military and logistical support from the United Kingdom, France, and the United States. The postwar failure to conduct a Nuremberg-style tribunal for Ustasha war crimes, combined with the Catholic Church's involvement in facilitating the escape of key perpetrators to Spain and Argentina, left unresolved grievances that later resurfaced in the ethnic conflicts between Serbs and Croats.

Following World War II, Josip Broz Tito's Communist regime sought to suppress ethnic antagonisms through the promotion of a unified, multinational state under the ideological banner of "Brotherhood and Unity" (Djilas and West, 1995). However, nationalist sentiments re-emerged in the early 1970s, prompting Tito to respond with a combination of political repression and purges. After Tito's death in 1980, the absence of a strong central authority capable of enforcing stabilization resulted in a power vacuum, with the collective leadership of the Communist Party failing to maintain cohesion (Petak, 2003). In this context, political elites in Serbia and Croatia increasingly embraced nationalist rhetoric, leveraging historical grievances to consolidate support and, ultimately, to mobilize their ethnic constituencies for armed conflict. Many of these leaders had previously been key figures within the socialist establishment but quickly transitioned from internationalist platforms to nationalist agendas, reframing economic hardships through the lens of ethnic victimization (Adamovich, 1995; Gligorov, 1996; Pošarac, 1996; Horowitz, 2003).

In addition to internal factors, external dynamics and the failure of international actors to provide effective third-party supervision contributed to the escalation of hostilities (Fearon, 1994). Ethnic polarization was further exacerbated by stark regional inequalities in wealth distribution and per capita income. These economic disparities had been widening since World War II (Kukić, 2020), fuelling inter-republic tensions and reinforcing nationalist narratives. By the late 1980s, the economic divergence between Yugoslav regions was striking: Kosovo, the least developed area, had a per capita income comparable to that of Pakistan, while Slovenia, the most developed republic, was on par with New Zealand (Borak, 2002). The unemployment rate ranged from as low as 3.2% in Slovenia's northwestern regions to a staggering 38% in Kosovo, highlighting the extreme regional imbalances that shaped economic and social grievances (Woodward, 1995).

The economic model of Yugoslavia had long been regarded as a grand experiment in combining the most feasible elements of capitalism and socialism, often referred to as a "third way" (Byrne, 2015; Rubinstein, 2015; Bianchini, 2020). The Yugoslav experiment attracted significant scholarly attention, with numerous economists analyzing its viability and long-term sustainability (Ward, 1967; Marschak, 1968; Bergson, 1987; Horvat, 1971; Furubotn, 1971; Sirc, 1979; Sapir, 1980; Gapinski et al., 1989; Estrin and Uvalic, 2008). However, despite its unique features, the model ultimately proved incapable of mitigating the centrifugal forces pulling the republics apart (Pejovich, 1990; Kukić, 2018). By the late 1980s, Yugoslavia's political and economic instability had reached a breaking point, paving the way for the violent disintegration of the federation.

# 4  Empirical Strategy

To rigorously assess the long-term impact of the Yugoslav civil war on regional economic growth, we employ a counterfactual approach that reconstructs the missing economic trajectory in the absence of conflict. Our strategy builds on the synthetic control estimator (Abadie et al., 2010, 2015), leveraging pre-conflict similarities in economic trajectories between the regions of the former Yugoslavia and a carefully selected donor pool of regions from 28 conflict-free countries. This allows us to isolate the war's causal effect on long-term development outcomes while mitigating concerns over confounding shocks.

Our empirical framework is grounded in the Rubin (1973) potential outcomes model with multiple treated units. Let $Y_{i,t}^I$ be the observed outcome and let $Y_{i,t}^N$ denote the outcome in the hypothetical absence of armed conflict that takes place at time $T_0$. Our primary objective is to estimate the treatment effect of armed conflict in former Yugoslavia, denoted as $\alpha_{1,t} = Y_{i,t}^I - Y_{i,t}^N$ for each post-conflict period $t > T_0$. To ensure robustness of the effect to unobserved time-varying heterogeneity, the observed regional per capita GDP is sought to serve as the best possible approximation of their counterfactual realization, satisfying both balancing conditions such that $Y_{i,t}^I = Y_{i,t}^N$ for $t < T_0$ to recover the full set of post-conflict outcome gaps, given by $\alpha = \{\alpha_{1,T_0+1}, \ldots, \alpha_{1,T}\}$ that per se does not depend on $t$.

To recover plausible counterfactual per capita GDP trajectories, we estimate a dynamic latent factor model (Doudchenko and Imbens, 2016; Ferman et al., 2020; Kaul et al., 2022) of the following form:

$$Y_{i,t}^N = \delta_t + \theta_t \mathbf{Z}_i + \lambda_t \mu_i + \epsilon_{i,t}$$

Let $\delta_t$ denote unobserved time-varying technology shocks to per capita GDP common to all regions, $\theta_t$ be a vector of unknown parameters, $\mathbf{Z}_i$ be the vector of observable pre-conflict benchmark levels of per capita GDP and the auxiliary covariates, $\lambda_t$ be a vector of unknown common factors, and $\mu_i$ be the vector of unknown factor loadings. The transitory shocks to per capita GDP are given by $\epsilon_{i,t}$ and are assumed to be identically and independently distributed (*i.i.d*). The key advantage of the dynamic factor model lies in its ability to capture heterogeneous responses to unobserved factors whilst embedding time trends into the counterfactual projection.

To construct the full set of artificial control units that best track and approximate pre-conflict, region-level per capita GDP trajectories, we define a vector of non-negative weights $W = (w_1, \ldots, w_J)$, restricted to be non-negative, convex, and additive, which implies that $w_j \geq 0$ for each $j \in \{2, \ldots, J+1\}$ and $\sum_{j=2}^{J+1} w_j = 1$.

To approximate the trajectory of the counterfactual scenario, $W$ can be estimated by finding the solution to the following single-nested minimization problem:

$$\hat{\mathbf{W}}(\hat{\mathbf{V}}) = \arg\min_{\mathbf{W} \in W} (\mathbf{Z}_1 - \mathbf{Z}_0 \mathbf{W})' \mathbf{V} (\mathbf{Z}_1 - \mathbf{Z}_0 \mathbf{W})$$

where $\mathbf{V}$ denotes the vector of pre-conflict benchmark per capita GDP levels and the auxiliary covariates. It should be stressed that $\mathbf{V}$ is chosen through a two-stage procedure. In the initial training period, $\hat{\mathbf{W}}(\hat{\mathbf{V}})$ is minimized by making use of Vanderbei's (1999) constrained quadratic optimization routine.[3] In the validation stage, the choice of $\hat{\mathbf{W}}(\mathbf{V})$ is cross-validated to minimize the out-of-sample prediction error through the following optimization problem:

$$\hat{\mathbf{V}} = \arg\min_{\mathbf{V} \in V} (\mathbf{Y}_1 - \mathbf{Y}_0 \mathbf{W}(\mathbf{V}))' (\mathbf{Y}_1 - \mathbf{Y}_0 \mathbf{W}(\mathbf{V}))$$

where $\mathbf{V}$ is a diagonal positive semi-definite invertible matrix of dimension $(K \times K)$ with $\text{tr}(\mathbf{V}) = 1$. More specifically, the matrix of $\mathbf{Y}_1$ is projected onto $\mathbf{Z}_1$ by imposing $v_k = \frac{|\beta_k|}{\sum_{k=1}^{K} |\beta_k|}$ where $v_k$ denotes the $k$-th diagonal element of $\mathbf{V}$ and $\beta_k$ is the $k$-th coefficient of the linear projection of $\mathbf{Y}_1$ on $\mathbf{Z}_1$. Each value of $\mathbf{W}$ represents a potential control unit as a convex combination of pre-conflict outcome- and covariate-implied attributes.

To obtain a plausible representation of regional growth and development trajectories in the (hypothetical) absence of armed conflicts, the artificial control groups are reweighted such that the conflicted-affected region $\mathbf{Z}_1$ will match the control group on at least some pre-conflict $\mathbf{Z}_{i,t}$. Unknown factor loadings are then matched automatically, and the projection of unobserved effects of the convexity locus is avoided.

---

[3]This routine uses a simple algorithm using interior point method to solve the quadratic programming problem under the imposed constraints. The implementation of the method takes place vis-á-vis C++ plugin where the standard tuning parameters are imposed such as 10% constraint for the tolerance of violation, the maximum number of iterations is set at 5,000 and the clipping bound for the variables is set to 10.

The counterfactual estimated per capita GDP trajectory is given by the error-minimizing counterfactual outcome that best reproduces the growth and development trajectories of the region:

$$Y_{W,t} = \sum_{j=2}^{J+1} w_j Y_{i,t} = \delta_t + \theta_t \left( \sum_{j=2}^{J+1} w_j \cdot Z_i \right) + \lambda_t \left( \sum_{j=2}^{J+1} w_i \cdot \mu_i \right) + \epsilon_{i,t}$$

where it is supposed that there exists $w$ such that the artificial control unit falls within the convex hull of the conflict-affected regions to achieve efficient matching in the pre-conflict period on the implicit attributes of the per capita GDP path and the additional covariates. Henceforth, an approximately unbiased estimator of the long-term effect of armed conflict is then given by: $\hat{\alpha}_{1,t} = Y_{i,t} - \sum_{j=2}^{J+1} w_j^* Y_{i,t}$.

Fitting $\{Y, Z\}$ is sufficient to match the counterfactual outcome with its observed counterpart in the pre-$T_0$ period, given that $\mu_i$ is unobservable. To ensure that the latent factor model provides a plausible representation of the counterfactual growth and development trajectory, we adopt the criteria proposed by Abadie (2021) and drop the observations that do not follow the factor model. This is achieved by properly restricting the donor pool to the regions from countries where armed conflict has not occurred, which ensures that the Stable Unit Treatment Value Assumption (SUTVA) is not violated, using mass causality and conflict intensity criteria from Brecke (1999) and the *Uppsala Conflict Data Program.*

To validate the credibility of our counterfactual estimates, we implement a series of robustness checks. Our framework accounts for a large number of treated units (Cavallo et al., 2013; Acemoglu et al., 2016; Gobillon and Magnac, 2016), ensuring robust treatment effect estimation. The Average Treatment Effect (ATE) and Average Treatment Effect on the Treated (ATET) parameters are derived by estimating separate vectors for each region and former republic, as follows:

$$\alpha = N^{-1} \sum_{i=1}^{N} \{\hat{\alpha}_{i,T_0+1}, \ldots, \hat{\alpha}_{i,T_0}\}$$

To quantify uncertainty, we invert the pre-conflict error parameter and construct empirical 95% confidence intervals, ensuring the precision of our estimates. Additionally, we rely on standard in-space treatment permutations to assess the uniqueness of treatment effects relative to placebo gaps (Ferman and Pinto, 2017; Hahn and Shi, 2017; Li, 2020). The placebo gaps are denoted as: $\alpha_{i \neq j}^{\text{Placebo}} = \{\alpha_{i,t} : j \neq 1\}$. Given the expectation of a negative treatment effect, the left-tailed p-value on the null hypothesis associated with the armed conflict is computed as:

$$\mathbb{P}_{(i,t>T_0)} = \Pr(|\alpha_{1,t}^{Placebo}| > |\alpha_{1,t}|) = \frac{\sum_{j \neq i} 1 \times (|\alpha_{1,t}^{Placebo}| > |\alpha_{1,t}|)}{J}$$

which captures the likelihood that the observed effect of armed conflict occurred by chance, highlighting effect uniqueness rather than statistical significance in the conventional sense. To enhance the reliability of our analysis, we exclude poorly fitting units through a two-step adjustment process, removing regions where pre-conflict prediction errors exceed twice the observed error. By default, stricter decision rules based on match-up quality parameters improve the robustness of our synthetic control estimates.

# 5 Data

## 5.1 Outcome variables, covariates and samples

Our analysis employs geographically granular dataset designed to investigate the effects of civil war on regional growth trajectories. Our outcome variable of interest is region-specific per capita GDP, expressed in constant 2005 international Geary–Khamis dollars, obtained from the large-scale and rich subnational dataset of Gennaioli et al. (2014). This measure serves as a consistent benchmark for evaluating the effect of armed conflicts on economic performance across regions over time.

To achieve model stability, we incorporate a comprehensive set of auxiliary covariates to capture the influence of potential confounders on the regional growth dynamics. These include past benchmark values of per capita GDP during the pre-conflict period, which capture initial economic conditions and growth dynamics before the onset of armed conflict. We also account for a rich set of physical geographic characteristics (Beck et al., 2018) such as mean annual temperature, altitude, latitude and longitude coordinates, precipitation levels and duration of sunshine that influence regional economic growth potential, latent indicators of the legal system (Chang et al., 2021), average years of education (Gennaioli et al., 2014) as a proxy for human capital, and population density (residents per square kilometer) as an conjugate indicator of regional demographic structure and cost of cooperation.

Our full sample consists of 43,559 region-year paired observations drawn from a multi-region, multi-country dataset spanning the period 1950–2015. This comprehensive dataset provides the necessary granularity to capture the overall effect of armed conflict on regional economic growth performance as well as the impact heterogeneity across regions and over time. The treatment sample focuses on 78 regions from the five former Yugoslav republics: Bosnia and Herzegovina, Croatia, North Macedonia, Serbia, and Slovenia.[4] The treatment sample includes 5,005 region-year paired observations, covering the period 1950–2015. These regions were directly affected by the Yugoslav civil war, allowing us to analyze the long-term economic consequences of armed conflict.

To construct a robust counterfactual for the treated sample of regions, we employ a donor pool, which consists of 593 regions from 32 countries that did not experience internal armed conflict during the same period (1950–2015).[5] The donor pool comprises 38,554 region-year paired observations, providing a diverse and stable set of control units which satisfies SUTVA assumption. By leveraging the donor pool against our treatment sample, we ensure that the counterfactual growth trajectories are constructed from regions that share similar pre-conflict characteristics with the treated regions but have not been affected by armed conflict, enhancing the credibility and validity of our results. By combining treated regions and a robust donor pool with detailed auxiliary covariates, our full sample provides the foundation for a rigorous empirical analysis of the long-term economic impacts of civil war in former Yugoslavia.

---

[4]Due to the lack of regional GDP per capita estimates, the data for Montenegro and Kosovo is not available for the same time period.

[5]Albania, Argentina, Australia, Austria, Belgium, Bolivia, Brazil, Canada, Chile, China, Colombia, Denmark, Finland, France, Germany, Greece, Hungary, Ireland, Italy, Japan, Mexico, Netherlands, Norway, Portugal, Spain, Sweden, Switzerland, Turkey, United Kingdom, United States, Uruguay, Venezuela.

## 5.2 Pre-conflict outcome and covariate-implied imbalance

Figure 1 presents region-level pre-conflict imbalance in the implied per capita GDP trajectories between former Yugoslav regions and its synthetic counterparts. Pre-conflict outcome and covariate-implied imbalance between the conflict-affected regions and their synthetic counterparts are very low. Relative to established benchmarks, the implicit imbalance is less than 1% (Adhikari and Alm, 2016), ensuring that the synthetic controls provide credible approximations of the counterfactual economic growth trajectories in the hypothetical absence of conflict. Notably, the inclusion of the full pre-conflict outcome path enhances the validity of the estimates without rendering auxiliary covariates irrelevant during the training stage (Kaul et al., 2022). This highlights the importance of balancing both observed outcomes and covariates to achieve accurate and reliable counterfactual projections.

It should be noted that our analysis incorporates a large number of synthetic control groups, leveraging Benford's law to approximate sparsity conditions and improve the reliability of matching procedures. The application of Benford's law ensures that overfitting is minimized, and the synthetic controls remain robust to variations in the underlying data structure (Rauch et al., 2014). Figure 2 depicts the frequency distribution of non-zero weight by donor region across different parts of former Yugoslavia. Distinct patterns emerge when examining the synthetic control composition for the former Yugoslav regions with notable contrasts and important similarities. Richer regions in the treatment sample, particularly in more advanced north-western part of former Yugoslavia (i.e. Slovenia) are most effectively synthesized by donor regions from advanced economies, including Germany, France, Japan, and Switzerland. These regions share similar pre-conflict growth trajectories and structural characteristics, making them well-suited for approximating the counterfactual scenarios of high-income regions. Conversely, poorer regions in the treatment sample (i.e. Bosnia and Herzegovina, Northern Macedonia) are synthesized through convex combinations of regions from emerging economies, such as Brazilian, Turkish, and Venezuelan states or provinces. These contrasts underscore the necessity of tailoring synthetic control construction to reflect the effect heterogeneity from the contextual setup.

In short, by ensuring a balanced comparison between treatment and control samples in pre-conflict period, low pre-conflict imbalances, and rigorous application of synthetic control methods, our study provides a robust foundation for assessing the long-term economic consequences of civil war.

# 6 Results

## 6.1 Baseline results

Our baseline findings, presented in Figure 3, reveal a substantial and persistent negative impact of the Yugoslav civil war on regional economic growth. Compared to the synthetic counterfactual scenario – constructed under the assumption of no armed conflict – we estimate an average decline of 38% in per capita GDP across the war-affected regions vis-á-vis the synthetic counterfactual. Contrary to standard theoretical predictions, which posit that the economic effects of civil war are temporary and gradually dissipate over time, we find that the negative shock induced by the conflict has remained largely permanent, with no evidence of full recovery up to the present. Figure 3 indicates no sign of pre-conflict anticipation for which we reject the null hypothesis (p-value = 0.000).

The economic consequences of armed conflict exhibit notable geographic variation across former Yugoslav regions (Figures 4 and 5). Less developed southeastern regions, particularly in North Macedonia and Serbia, suffered the most severe economic contractions, with prolonged stagnation and little sign of convergence to pre-conflict growth trajectories of their synthetic peers. By contrast, northwestern regions—notably those in closer proximity to Austria and Italy, such as Slovenia—experienced significantly less economic disruption and tend to experience a more transitory negative per capita GDP effect of war. The ability of these regions to integrate into Western European institutional integration, coupled with stronger institutional resilience, likely mitigated the long-term economic costs of the war. Moreover, capital city regions appear to have been significantly more resilient than non-capital regions. The estimated negative effect on capital cities is roughly half the magnitude of that observed in non-capital regions, reflecting the concentration of post-conflict reconstruction efforts, foreign investment, and institutional stability in urban centers. Additionally, the economic damage appears to dissipate more rapidly in wealthier regions, particularly those in closer geographic and cultural proximity to Western Europe, where the per capita GDP gap is approximately 60% lower relative to more isolated regions.

Figure 4 reports region-specific average per capita GDP decline for four republics in former Yugoslavia. For clarity, Figure 5 separately shows the corresponding estimates for Slovenian regions. The spatial disparity in the magnitude of the average per capita GDP decline in response to the armed conflict is substantial. For instance, while only nine regions – or roughly 11 percent of the treatment sample – tend to experience a temporary deviation of the growth trajectory with no long-term repercussions – the estimated per capita GDP decline appears to be permanent and statistically significant across 89 percent of the treatment sample. The sub-national effect is thus characterized by the significant permanent derailment of the economic growth trajectory vis-á-vis the synthetic counterfactual. Republic-specific evidence and estimates are more exhaustively reported in the Supplementary Appendix.

## 6.2 Robustness checks and placebo analysis

**Large-sample permutation.** To assess the significance of our estimates, we employ a large-sample permutation inference framework, adapting the methodology of Doudchenko and Imbens (2016) to account for the complexities of non-parametric estimation, where inference under the sharp null hypothesis is rarely feasible. Given the presence of a large number of simultaneously treated units (N = 348), we generate in-space placebo distributions by permuting the armed conflict to the unaffected regions in the donor pool whilst shifting the Yugoslav regions from the treatment sample into the donor pool. This approach enables us to compute a Fisher-exact p-value by comparing the real treatment effect with the distribution of placebo estimates, using the gap-level RMSE ratio as the primary test statistic. Unlike standard permutation tests, our full-random sampling permutation algorithm ensures that treatment assignment remains statistically independent across simulations, enhancing robustness. To determine the permanence or transience of the estimated effects, we evaluate whether each treated entity or group of entities exhibits a sustained deviation from its synthetic counterfactual. The scale of our inference procedure is substantial: with 12.5 billion placebo simulations, we provide a more precise assessment of statistical significance. The large-N framework further ensures that asymptotic stability remains intact, mitigating biases between treated units and their control groups and reinforcing the validity of our estimates. Figure 6 reports the intertemporal distribution of p-values on the null hypothesis of no effect whatsoever across country-level subsamples. The evidence almost unequivocally suggests that the war is associated with statistically significant downward derailment of per capita GDP trajectory that does not disappear up to the end-of-sample period.

**Interactive fixed-effects.** The estimated per capita GDP gap between conflict-affected regions and their synthetic counterparts may be subject to uncertainty arising from migration, policy responses, and capital flight, among other post-conflict economic dynamics. To rigorously assess the statistical significance of our estimates, we employ an interactive fixed-effects algorithm (Xu, 2017) and re-estimate the treatment effect of armed conflict in former Yugoslavia using a generalized synthetic control approach. We construct empirical confidence intervals by leveraging a placebo test distribution to reassign the armed conflict treatment to unaffected regions, shifting the Yugoslav regions from the treatment group to the donor pool. This process generates a large set of pseudo-treatment effects, forming a reference-based placebo distribution from which we derive two-tailed confidence intervals as the primary measure of statistical significance. Our inference is based on 1,000 bootstrap replications across the full sample and each sub-sample.

The estimated 38% decline in per capita GDP remains robust across alternative estimation techniques, including the matrix completion (Athey et al., 2021) and synthetic difference-in-differences (Arkhangelsky et al., 2021) estimates reported in Table 1. The latter estimates the treatment effect by matching and reweighing pre-intervention trends to weaken the reliance on parallel trend assumption, invariance to the additive region-level shifts can yield more efficient estimates compared to synthetic control method.[6] Three insights emerge from a hybrid difference-in-differences and synthetic control estimation. First, the per capita GDP decline is sizeable and appears to be permanent relative to the unit- and time-contingent control groups. Second, the estimated decrease in per capita GDP vis-á-vis the control group with non-additive shifts is -40%. And third, richer regions in the northwest tend to have around 60% smaller effect magnitude. In absolute terms, the estimated 40 percent decline translates to a $6,000 G-K drop in per capita GDP, an economic contraction nearly three times the magnitude of the Palestinian Intifada's impact on the Israeli economy (2000) identified by (Horiuchi and Mayerson, 2015). The estimated per capita GDP remains statistically significant across all specifications, as depicted in Figure 7, which reports the two-tailed 95% empirical confidence intervals. While controlling for parallel trends in the pre-conflict period attenuates the estimated GDP decline by approximately 20%, the effect persists over time with no signs of reversal, underscoring the permanent economic impact induced by the war.

**In-time placebo**. To further assess the robustness of our estimates, we implement an in-time placebo analysis, in which the timing of the conflict is falsely reassigned to a pre-treatment sub-period within pre-conflict training and validation period, during which the estimated treatment effect should be indistinguishable from zero if the inference on the effect is valid. This approach allows us to test whether the observed per capita GDP gap emerges only after the actual onset of conflict, rather than being driven by pre-existing trends or model artefacts. The results, reported in Figure 8, provide strong evidence that the null hypothesis of in-time placebo gap equality can be consistently rejected at the conventional 5% significance level, reinforcing the credibility of our estimated treatment effect and ruling out spurious attribution of economic divergence in the treated regions prior to the outbreak of the conflict.

---

[6]In Supplementary Appendix S1, we re-estimate the counterfactual scenario through hybrid difference-in-differences and synthetic control method (Arkhangelsky et al., 2021). By matching and reweighing pre-conflict per capita GDP trends to weaken the reliance on parallel trend assumption, synthetic difference-in-differences method is particularly invariant to the additive region-level shifts in comparison with the latent factor model commonly used in the counterfactual scenario estimation, and provides flexible parametric inference on the treatment effect underscored by a sharp null hypothesis. The correlation between our baseline synthetic control estimates and synthetic difference-in-differences estimates is between +0.8 and +0.9 across country-level subsamples and is statistically significant at 1% (i.e. p-value = 0.000). Synthetic difference-in-differences estimated are exhibited in Supplementary Appendix S1.

**Matching on salient economic features.** A critical methodological concern in estimating the long-term effects of armed conflicts using the synthetic control method is the selection of an appropriate donor pool for the analysis. A worldwide donor pool consisting of regions without any armed internal conflict in both the pre- and post-conflict period is ought to be evaluated in terms of its ability to capture the key characteristics of the pre-1987 Yugoslav economy, and is prone to overfitting if the number of potential donor regions is large. Ideally, a well-constructed donor pool should reflect institutional, geographic, cultural, historical, and economic development attributes as closely as possible to those of Yugoslavia from 1950 onward.

In particular, it is pivotal that the donor pool accounts for both geographic proximity and the presence of non-democratic regimes similar to the Yugoslav one, as these factors play a critical role in shaping economic and institutional trajectories. Previous research has established that democratic regimes tend to exhibit distinct growth trajectories compared to non-democracies (Acemoglu et al., 2019). Consequently, if the counterfactual donor pool includes predominantly democratic countries such as Switzerland and France with uninterrupted political liberalization since 1950 onwards, the estimated impact of armed conflict could become partially conflated with the effects of political liberalization, potentially biasing the results (Giavazzi and Tabellini, 2005).

To mitigate this risk, a more refined approach involves restricting the donor pool to a regional-level sample from Southern European countries and Türkiye. This selection is motivated by the post-war experience of certain countries – Greece, Portugal, Spain, and Türkiye – which maintained non-democratic regimes and low degree of economic liberalization for extended periods, as well as Italy, which shared deep historical and geographic ties with Yugoslavia. Such more compact and less noisy donor pool consists of 165 regions across five Mediterranean European countries[7], covering the period 1950–2015, resulting in 10,890 region-paired observations. By adjusting the donor pool in this way, our analysis reduces the risk of overfitting, ensuring that the latent factor model used in the estimation process remains robust and well-calibrated for capturing the causal effects of war while minimizing bias arising from political regime-type heterogeneity.

Figure 9 presents the full-sample per capita GDP effect of armed conflict based on salient southern European features of the Yugoslav economy in the 1950-1987 period. Using a Mediterranean region-level donor pool from Italy and non-democratic regimes (1950–1980s), our estimates indicate a 46% decline in per capita GDP, ceteris paribus. The estimated magnitude of decline remains highly correlated with our baseline estimates (+0.91, p-value = 0.000), reinforcing the robustness of our findings. While some degree of economic recovery is observed following the Dayton Agreement (1995), the overall decline relative to the condensed synthetic control groups appears to be permanent, with no statistical evidence of a temporary war effect persisting through 2015. The long-term economic loss and the breakdown of the per capita GDP trajectory are further reflected in significant heterogeneity both across and within former Yugoslav republics. This suggests that regional disparities in the impact of war were shaped by underlying structural and institutional factors. These findings underscore the enduring nature of war-induced economic disruptions and challenge narratives that emphasize short-lived macroeconomic consequences of civil conflict. Figure 10 depicts country-level disparities in the magnitude of the effect, uncovering non-trivial heterogeneity of the impact.

Although the Mediterranean donor provides an excellent quality of the fit in reproducing pre-conflict growth trajectories of Yugoslav regions, not all Mediterranean regions synthesize and reproduce the respective trajectories on equal footing. Figure 11 presents frequency of non-zero weight for the Mediterranean donor regions for the full sample of conflict-affected Yugoslav regions. For instance, the per capita GDP trajectories of the least developed regions prior to the outbreak of the conflict are best synthesized by a convex combination of the implied attributes of predominantly Turkish regions. By way of example, the per capita GDP trajectory of the Herzegovina-Neretva Canton prior to the outbreak of the conflict is best synthesized by the convex combination of the implicit attributes of Agri (70%), Adiyaman (20%), Alto Adige (5%), Calabria (3%), and Anatoliki Makedonia and Thraki (2%), respectively. In more advanced regions, the synthetic control groups are dominated by the density of more advanced regions from Türkiye, Italy and Greece. By contrast, the per capita GDP trajectory of the most advanced and industrialized regions before the outbreak of the conflict are best synthesized by the convex combination of the implied characteristics of the more advanced Italian and Spanish regions. For instance, the synthetic versions of Upper Carniola and Ljubljana-Central Slovenia load strongly on the latent characteristics of Madrid, Tarragona, Friuli Venezia Giulia, Álava, Teruel, Istanbul and several others. In both cases, the evidence uncovers a temporary deviation of the growth trajectory followed by the outperformance of the synthetic counterfactual scenario (i.e. Ljubljana-Central Slovenia) in contrast to the structural long-term breakdown of the per capita GDP trajectory (i.e. Upper Carniola). Supplementary Appendix S2 provides a more elaborate analysis of the treatment effect using the Mediterranean donor pool.

---

[7]Greece, Italy, Portugal, Spain, Türkiye.

### 6.3 Mechanisms

A key question remains: through which mechanisms does armed conflict translate into regional economic losses? The literature suggests many potential channels (Section 2). We are able to measure several of them with regional data, including population displacement, ethnic fractionalization and polarization, geographical factors, and historical legacies of imperial rule. We collected regional data on population size and density as well as ethnic composition (from a series of Statistical Yearbooks of Yugoslavia), geographical characteristics such as latitude and distance to coast (Gennaioli et al., 2014), and years under Habsburg and Ottoman rule (Brown, 1996; Dimitrova-Grajzl, 2007).

Table 2 reports the identified correlates and channels behind the relatively large regional per capita GDP gap variance. Each specification contains the full set of republic-fixed effects to account for potentially unobserved cross-regional gap heterogeneity. We find that forced migration and population displacement account for the largest share of cross-regional per capita GDP variance. Specifically, we find that a 1 percent increase in displacement intensity – proxied by changes in population density before and after the war – is associated with a 2 percent per capita GDP loss vis-á-vis the synthetic counterfactual, ceteris paribus, with displacement explaining approximately one-third of the total economic decline.[8] Furthermore, ethnic conflicts and polarization account for around 40 percent of the observed GDP losses.[9] Regions with deeply fragmented and polarized social structures experience substantially greater magnitude of economic decline due to weakened intergroup cooperation in the post-conflict period. Geographic proximity to Western and Central Europe appears to mildly mitigate losses, suggesting that access to external markets and Western Europe in post-conflict stabilization are important drivers of recovery. Contrary to traditional historical narratives, regions formerly under Habsburg rule do not exhibit significantly lower economic losses (Becker et al., 2016), while there is some evidence that Ottoman historical legacy is associated with more severe per capita GDP losses (Pamuk, 2024). These findings highlight the multidimensional nature of the economic impact of war, where structural, demographic, geographical, and institutional factors interact to shape the long-term regional divergence.

# 7 Conclusion

The disintegration of former Yugoslavia was preceded by a prolonged institutional crisis after Tito's death in 1980, which fueled Serbo-Croatian ethnic nationalism and set the stage for a decade-long civil war. The armed conflict between 1988 and 1995 resulted in devastating humanitarian losses and inflicted irreversible economic damage across former Yugoslavia. Using a novel, regional dataset of per capita GDP (2005 Geary-Khamis dollars) covering a wide set of regions from former Yugoslavia and beyond, this study provides a counterfactual estimate of the regional economic cost of the Yugoslav war by applying synthetic control and difference-in-differences methods. Our findings indicate that the decade-long civil war led to a 38 percent decline in per capita GDP, permanently derailing the trajectory of economic growth as compared with the regional synthetic counterparts. These findings demonstrate the severe and persistent nature of war-induced economic damage.

---

[8] The share of per capita GDP decline behind the synthetic counterfactual explained by the population displacement is estimated based on the specifications (1) and (2) from Table 2 by parsing out the republic fixed-effects.

[9] The percentage of explained losses is estimated by looking at the share of overall variance in per capita GDP gaps vis-á-vis the synthetic counterfactuals jointly accounted for the ethnic diversity, concentation and polarization measures.

The primary drivers of these economic losses are long-term population displacement and deep-rooted Serbo-Croat ethnic hostilities and tensions, which created lasting structural and institutional disruptions to the Yugoslav economy and its successor states. Furthermore, while the north-western, wealthier regions of Yugoslavia exhibit a temporary economic effect of war, the impact elsewhere remains permanent and deeply detrimental.

Taken together, our findings highlight that the civil war in former Yugoslavia constituted an economic shock of extraordinary magnitude, with long-term deterioration in regional growth trajectories. Beyond the regional context, these results contribute to the broader literature the long-term economic consequences of war and post-conflict recovery, emphasizing the importance of population displacement and ethnic divisions in conditioning economic decline. Future research should explore the role of international interventions, economic integration, and migration policies in shaping divergent post-war recovery paths.

# Figures

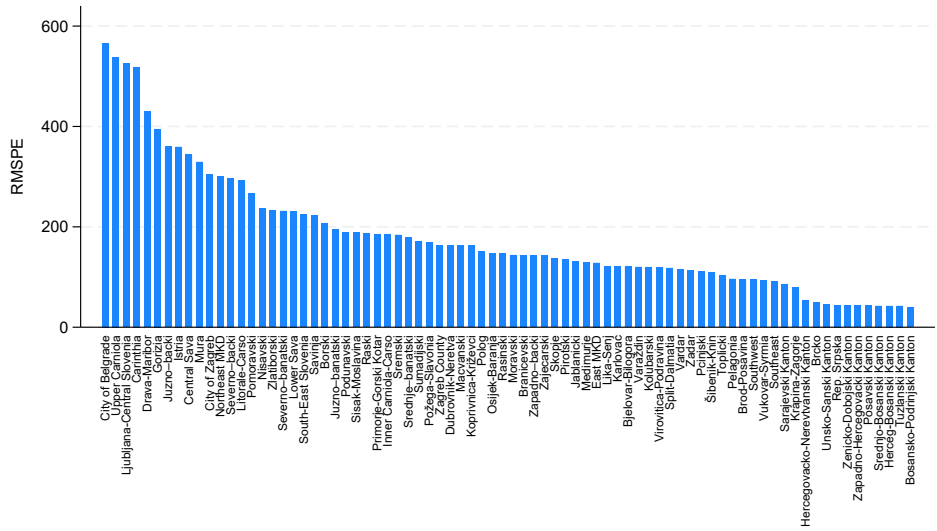

**Figure 1:** *Pre-conflict region-level imbalance of per capita GDP trajectories*

**Figure 2:** *Composition of synthetic control groups using sparse Benford frequency distribution*

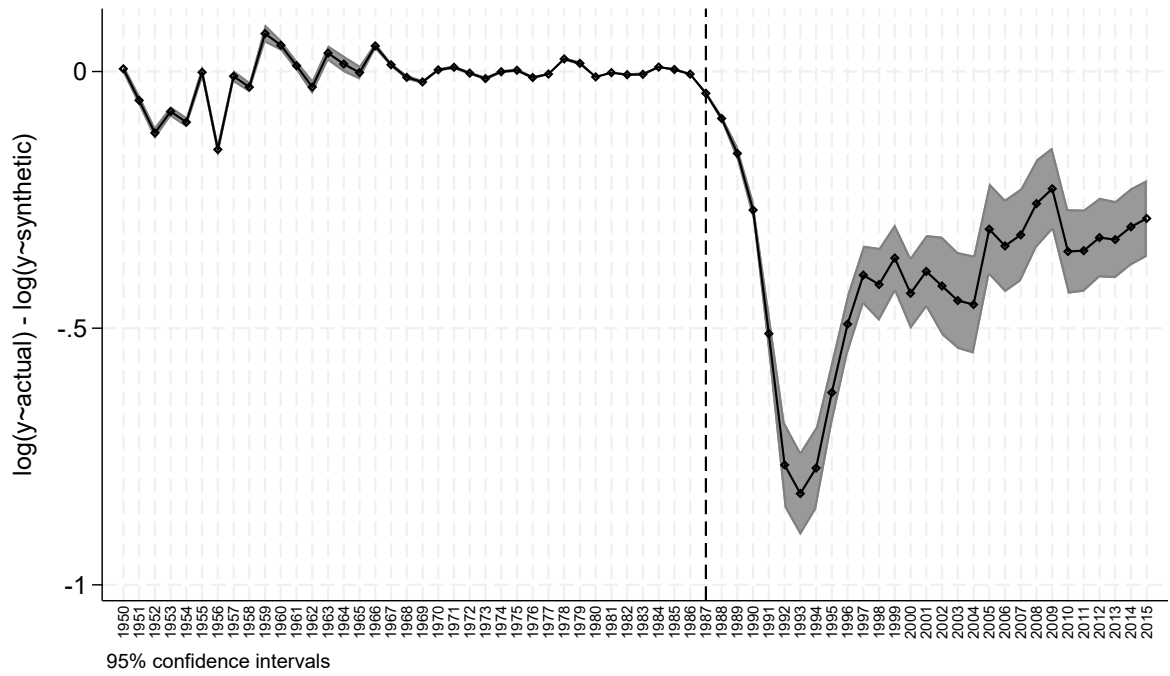

**Figure 3:** *Long-term effect of armed conflict in former Yugoslavia on economic growth across full treatment sample, 1950-2015*

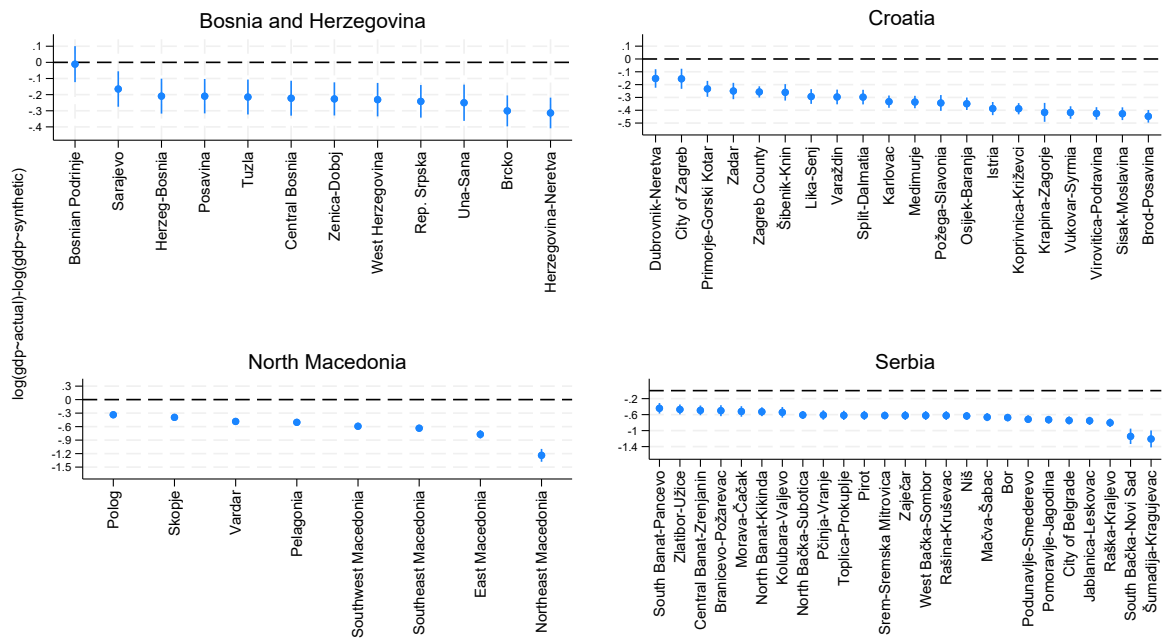

**Figure 4:** *Long-term effect of Yugoslav war on economic growth trajectories of regions within former Yugoslav republics, 1950-2015*

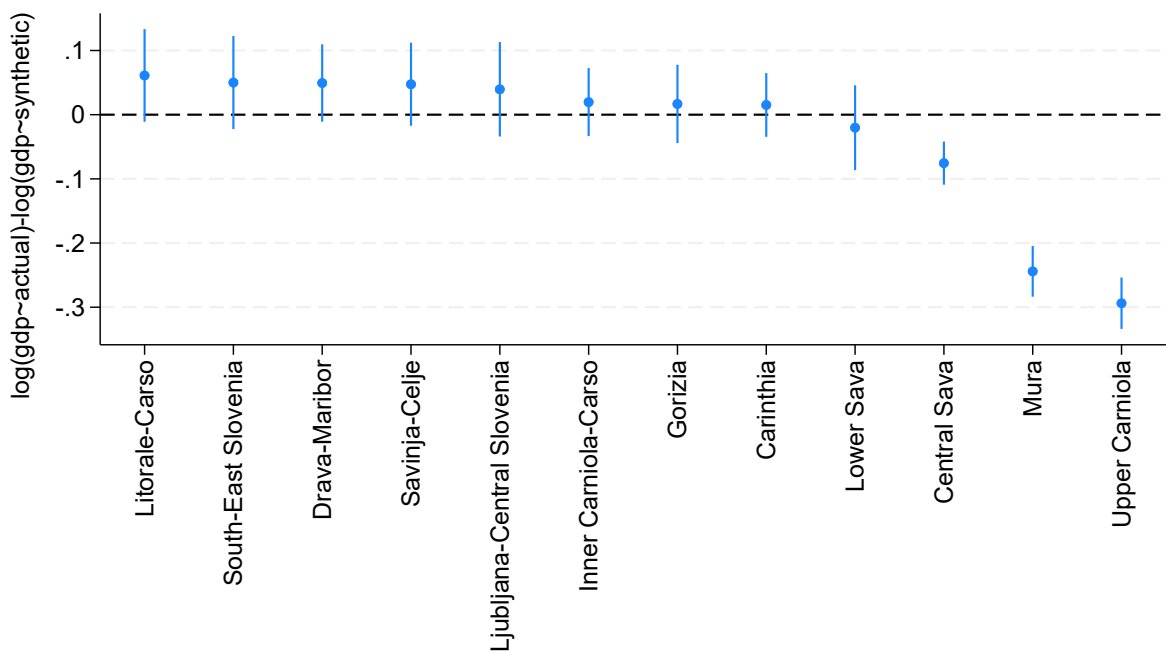

**Figure 5:** *Long-term effect of Yugoslav war on economic growth trajectories of Slovenian regions, 1950-2015*

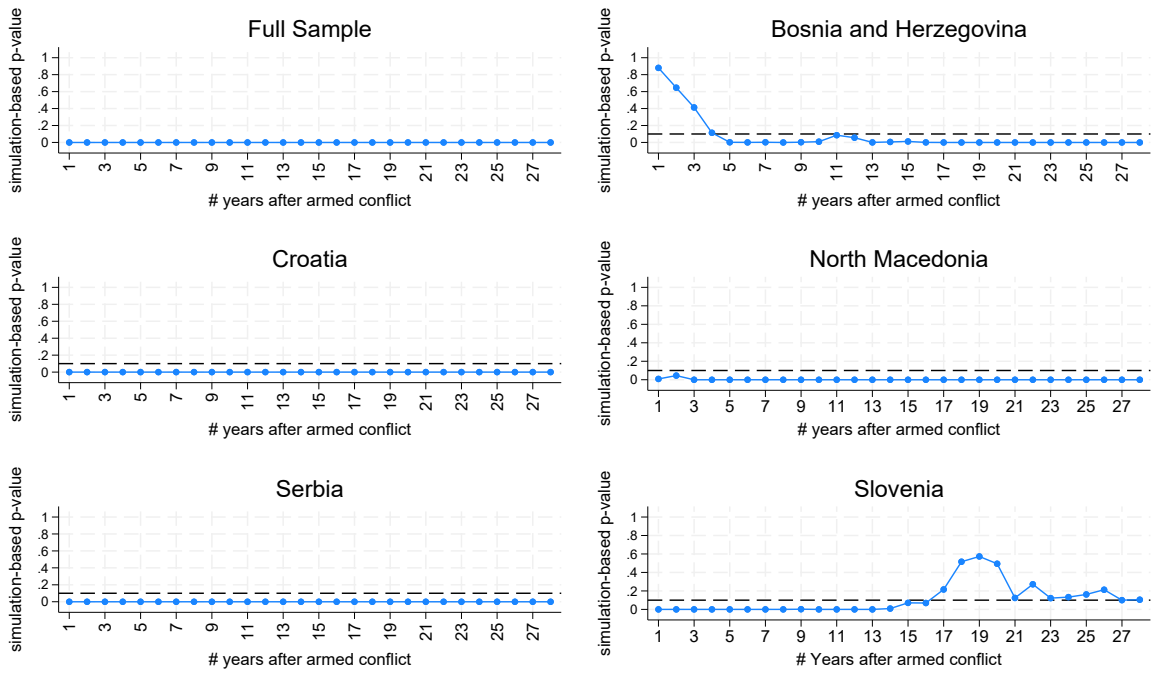

**Figure 6:** *Inference on the long-term subnational growth impact of civil war in former Yugoslavia, 1950-2015*

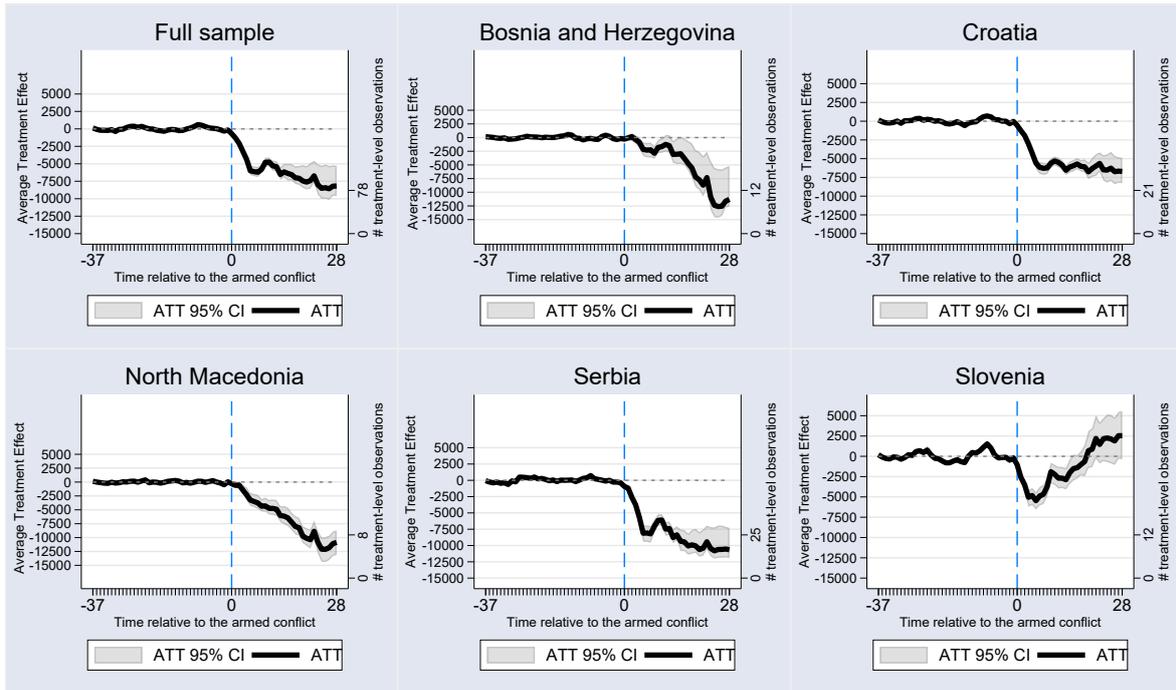

**Figure 7:** *Long-term impact of civil war on subnational economic growth trajectories across former Yugoslavia using interactive fixed-effects algorithm, 1950-2015*

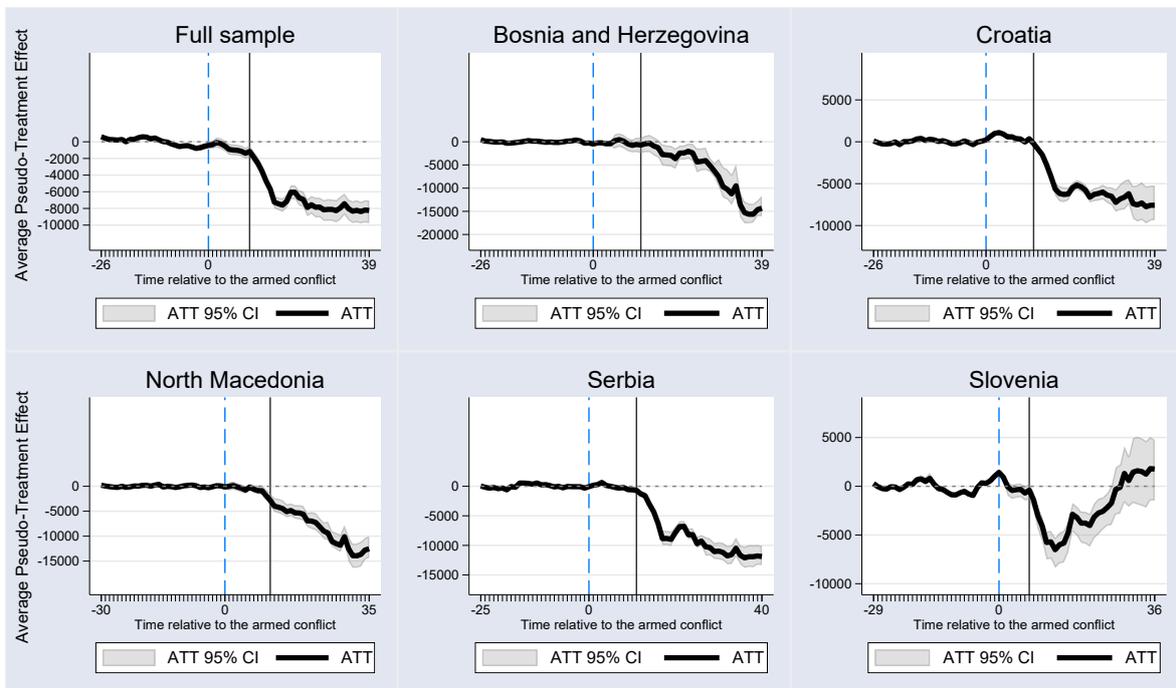

**Figure 8:** *In-time placebo analysis of the long-term economic growth impact of civil war in former Yugoslavia, 1950-2015*

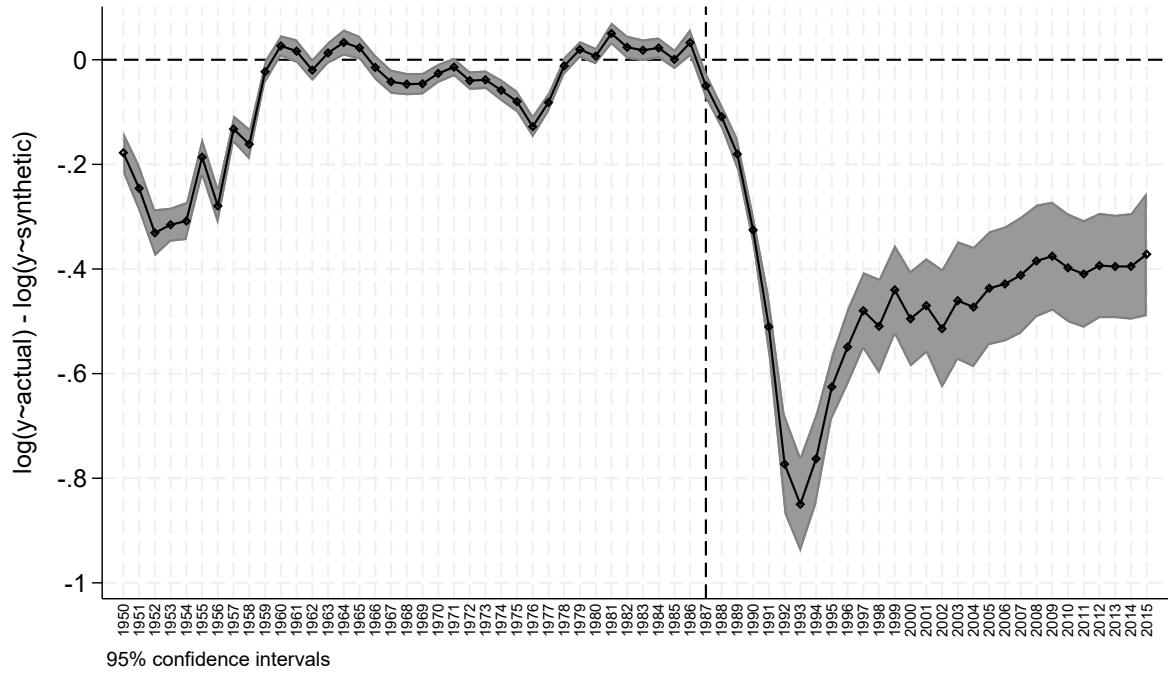

**Figure 9:** *Re-estimated long-term growth effect of civil war using Mediterranean region-level donor pool, 1950-2015*

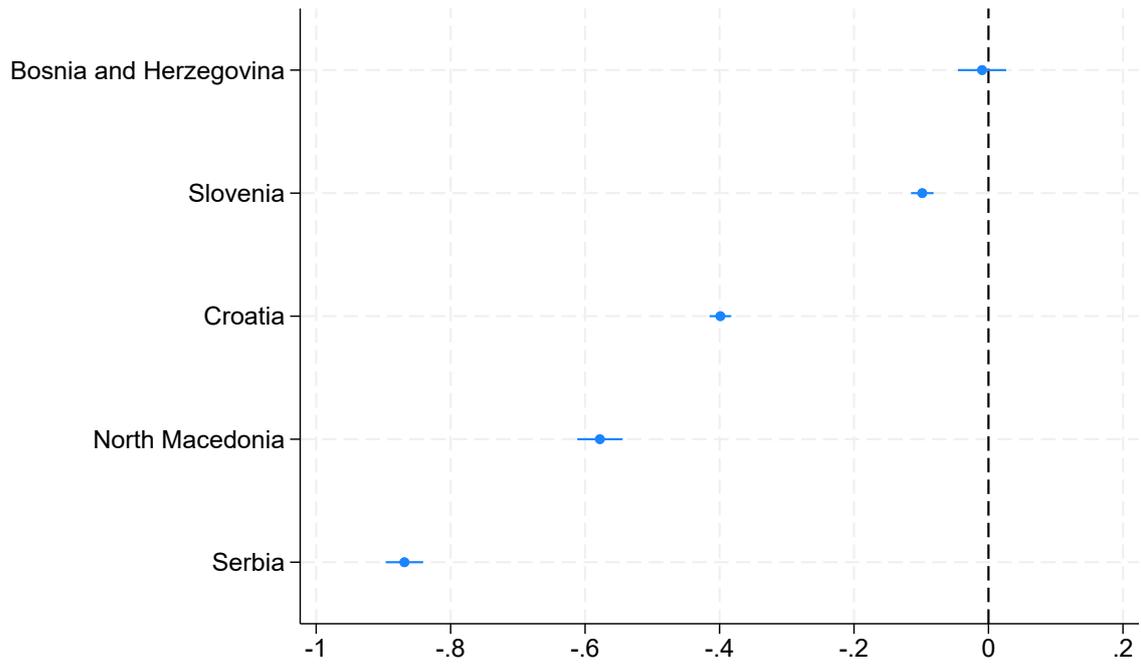

**Figure 10:** *Country-level average per capita GDP effect of civil war using Mediterranean donor pool, 1987-2015*

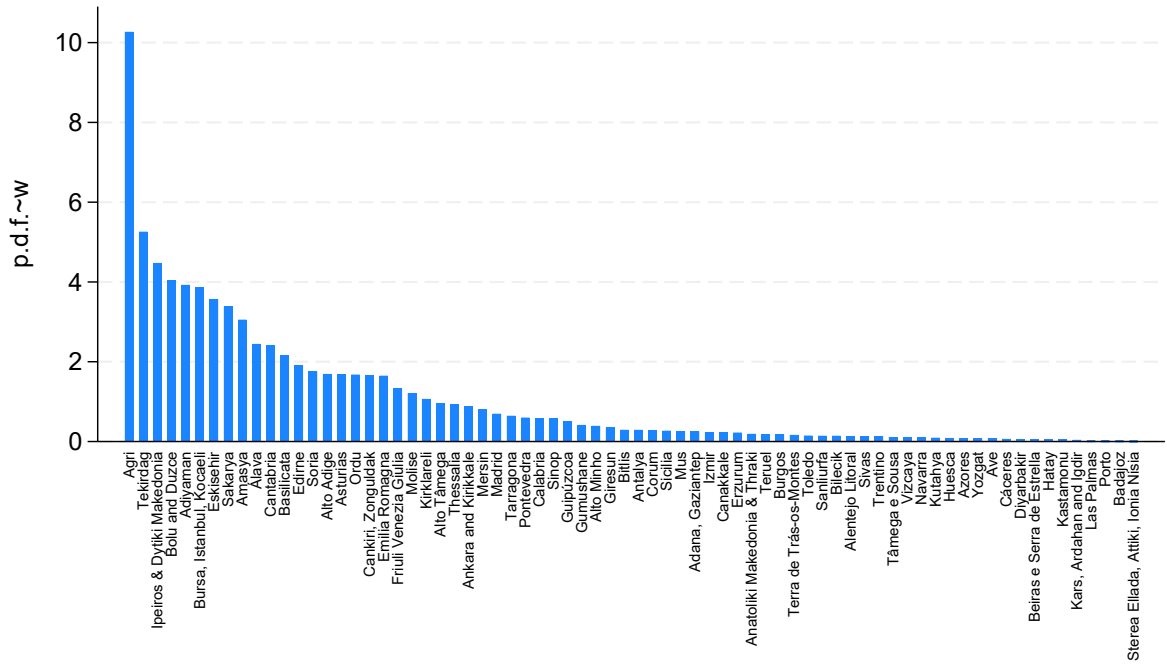

**Figure 11:** *Density of non-zero weight frequency across Mediterranean donor regions, 1950-2015*

# Tables

**Table 1:** *Interactive fixed-effects and matrix policy completion estimates, 1950-2015*

|  | Full Sample | Analysis by Sub-Sample | | | | |
|---|---|---|---|---|---|---|
|  |  | Bosnia and Herzegovina | Croatia | North Macedonia | Serbia | Slovenia |
|  | (1) | (2) | (3) | (4) | (5) | (6) |
| **Panel A: Fixed-effects difference-in-differences estimates** | | | | | | |
| Post-$T_0 \Delta Y$ | -6895.62 | -7924.69 | -6214.54 | -8643.61 | -9354.99 | -1193.24 |
|  | (416.76) | (386.85) | (433.87) | (463.18) | (390.48) | (780.47) |
| Two-tailed 95% confidence interval | (-7787.24, -6082.21) | (-8583.66, -7176.51) | (-7032.85, -5353.51) | (-9617.59, -7664.55) | (-10086.66, -8591.54) | (-2612.61, 390.33) |
| Simulation-based p-value | 0.000 | 0.000 | 0.000 | 0.000 | 0.000 | 0.145 |
| **Panel B: Interactive fixed-effects generalized synthetic control estimates** | | | | | | |
| Post-$T_0 \Delta Y$ | -6215.52 | -5080.94 | -5623.39 | -6729.09 | -8278.55 | -1247.17 |
|  | (524.46) | (865.74) | (310.51) | (500.73) | (622.16) | (699.47) |
| Two-tailed 95% confidence interval | (-6817.04, -4836.51) | (-5853.34, -1634.78) | (-6146.23, -4916.79) | (-7664.84, -5685.21) | (-9081.80, -6595.34) | (-2538.85, 55.205) |
| Simulation-based p-value | 0.000 | 0.000 | 0.000 | 0.000 | 0.000 | 0.075 |
| **Panel C: Matrix completion algorithm** | | | | | | |
| Post-$T_0 \Delta Y$ | -6895.59 | -7859.38 | -6149.32 | -8578.31 | -9289.75 | -1128.07 |
|  | (409.55) | (408.43) | (449.94) | (460.23) | (378.09) | (828.08) |
| Two-tailed 95% confidence interval | (-7581.67, -6022.48) | (-8580.52, -6990.09) | (-6911.72, -5082.20) | (-9455.59, -7526.10) | (-10049.34, -8659.71) | (-2645.56, 488.72) |
| Simulation-based p-value | 0.000 | 0.000 | 0.000 | 0.000 | 0.000 | 0.000 |
| **Panel D: Synthetic difference-in-differences** | | | | | | |
| Post-$T_0 \Delta Y$ | -4613.01 | -7832.17 | -5350.37 | -4390.32 | -6573.31 | -2403.03 |
|  | (355.86) | (178.66) | (518.04) | (594.17) | (640.31) | (685.55) |
| Two-tailed 95% confidence interval | (-5301.07, -3914.83) | (-11303.32, -4330.21) | (-6373.34, -4334.35) | (-5563.41, -3230.43) | (-7821.31, -5310.33) | (-3753.31, -1063.34) |
| Large-sample p-value approximation | 0.000 | 0.000 | 0.000 | 0.000 | 0.000 | 0.000 |

*Notes:* The table reports synthetic difference-in-differences estimated effects of the civil war on subnational per capita GDP trajectories of former Yugoslavia by matching regional pre-war economic growth trajectories with a worldwide regional donor pool without any armed internal conflict. It reports the weighted difference between former Yugoslav region-level growth trajectories and their synthetic peers based on the localized two-way fixed effect estimator. The outcome model includes latent region-level factors interacted with latent time factors. It also reports latent time-varying weights and large sample-approximated empirical p-values on the null hypothesis behind the average treatment effect. The lower and upper bounds of the two-tailed 95% confidence intervals are reported in parentheses.

**Table 2:** *Correlates of per capita GDP losses from Yugoslav war across the full sample of regions, 1988-2015*

| | Population Displacement | | Ethnic Fractionalization and Polarization | | | | Economic Geography | | Historical Persistence | | All |
|---|---|---|---|---|---|---|---|---|---|---|---|
| | (1) | (2) | (3) | (4) | (5) | (6) | (7) | (8) | (9) | (10) | (11) |
| **Panel A: Population Displacement** | | | | | | | | | | | |
| $\ln(\text{Pop.Density})_{t<T_0} - \ln(\text{Pop.Density})_{t<T_0}$ | -0.213*** | | | | | | | | | | -0.594** |
| | (0.021) | | | | | | | | | | (0.289) |
| $\ln(\text{Pop.Size})_{t<T_0} - \ln(\text{Pop.Size})_{t<T_0}$ | | 0.001 | | | | | | | | | 0.004 |
| | | (0.073) | | | | | | | | | (0.044) |
| **Panel B: Ethnic Fractionalization and Polarization** | | | | | | | | | | | |
| Ethnic Diversity (Alesina-Ferrara) | | | -0.227*** | | | | | | | | 0.481 |
| | | | (0.053) | | | | | | | | (1.749) |
| Ethnic Concentration (HHI) | | | | -0.224*** | | | | | | | 0.332 |
| | | | | (0.066) | | | | | | | (1.739) |
| Ethnic Polarization (Reynal-Querol) | | | | | -0.173*** | | | | | | -0.157 |
| | | | | | (0.039) | | | | | | (0.139) |
| Fraction Serbs | | | | | | -0.355*** | | | | | -0.160 |
| | | | | | | (0.117) | | | | | (0.124) |
| **Panel C: Economic Geography** | | | | | | | | | | | |
| Latitude | | | | | | | -0.0004** | | | | 0.003 |
| | | | | | | | (0.00004) | | | | (0.035) |
| Inverse Distance from Coast | | | | | | | | 0.299 | | | 0.777 |
| | | | | | | | | (0.341) | | | (0.564) |
| **Panel D: Historical Persistence** | | | | | | | | | | | |
| Years under Habsburg Rule | | | | | | | | | -0.0001 | | -0.0002 |
| | | | | | | | | | (0.0002) | | (0.0002) |
| Years under Ottoman Rule | | | | | | | | | | -0.0004*** | -0.00006 |
| | | | | | | | | | | (0.00007) | (0.0003) |
| Republic-Fixed Effects | YES | YES | YES | YES | YES | YES | YES | YES | YES | YES | YES |
| Observations | 78 | 78 | 78 | 78 | 78 | 78 | 78 | 78 | 77 | 77 | 77 |
| $R^2$ | 0.90 | 0.87 | 0.90 | 0.90 | 0.82 | 0.88 | 0.90 | 0.87 | 0.90 | 0.91 | 0.93 |

*Notes:* Cluster-robust standard errors are denoted in parentheses. Asterisks denote statistically significant coefficients at 10% (*), 5% (**), and 1% (***), respectively.

# Supplementary Appendix

## S.1 Republic-specific region-level evidence

### S.1.1 Bosnia and Herzegovina

Our subnational evidence for Bosnia and Herzegovina reveals an average per capita GDP decline of approximately 22 percent in response to the armed conflict (p-value = 0.000). In contrast to other Yugoslav republics, however, the economic consequences of civil war in Bosnia and Herzegovina appear to be relatively more transient. We posit that substantial humanitarian and development aid played a pivotal role in facilitating rapid post-war economic recovery, enabling numerous regions to quickly return to their pre-war growth trajectories.

Figure 12 illustrates the estimated per capita GDP decline for Sarajevo and Herzegovina-Neretva Canton in comparison to their synthetic counterfactuals. The comparison of the effect magnitude over time reveal a stark contrast. Sarajevo experienced a significant, temporary deviation from its pre-conflict GDP per capita path, followed by a swift recovery. In contrast, Herzegovina-Neretva Canton faced a persistent and more severe breakdown in its economic growth trajectory. Specifically, our synthetic control estimates show that Sarajevo's per capita GDP dropped by nearly two-thirds by the end of the war, but swiftly rebounded, converging toward the trajectory implied by its synthetic control group. In stark contrast, Herzegovina-Neretva Canton saw an average per capita GDP decline of 31 percent, with no discernible temporary deviation, and by the end of the sample period, its per capita GDP remained 34 percent lower than that of its synthetic counterpart, which had not experienced conflict in the pre- and post-intervention period.

The synthetic control groups for Bosnia are primarily composed of regions such as Sichuan and Chongqing in China, Piaui in Brazil, and various others, including Venezuelan and Turkish regions. For example, Sarajevo's pre-war per capita GDP trajectory is most accurately reproduced as a weighted combination of Sichuan and Chongqing (46%), Piaui (29%), Yucatan (7%), Sergipe (5%), Wakayama (3%), Tabasco (3%), and several smaller contributors. Meanwhile, Herzegovina-Neretva Canton's pre-war trajectory is best approximated by a different combination of these same regions, though with a distinct weight distribution.

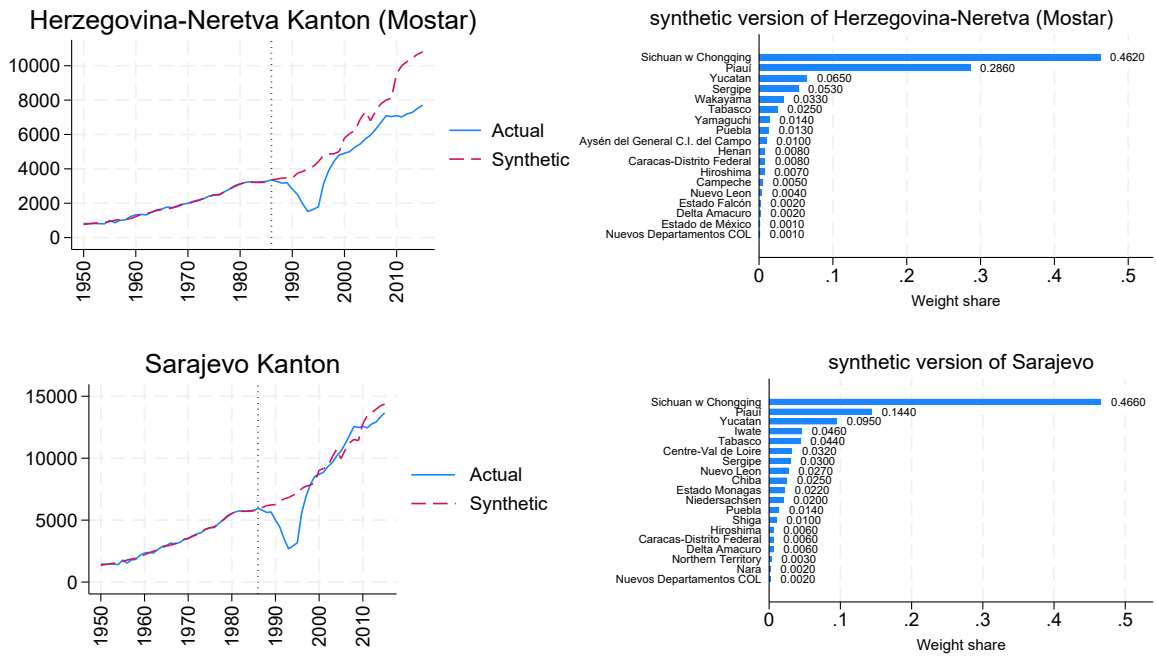

**Figure 12:** *Long-term effect of Yugoslav war on economic growth trajectories of Bosnian cantons, 1950-2015*

### S.1.2 Croatia

Our subnational evidence for Croatia indicates that the average per capita GDP decline in response to the armed conflict is around 32 percent (i.e. p-value = 0.000). Against this backdrop, economic effects of civil war appear to far less temporary compared to Bosnia and Herzegovina Figure 13 exhibits the estimated per capita GDP decline in response to the armed conflict for Zagreb City (i.e. the least adversely affected region) and Slavonski Brod-Posavina (i.e. most adversely affected region) vis-á-vis their synthetic counterfactuals. The estimated gaps unciver a large-scale negative temporary deviation of the per capita GDP trajectory in Zagreb against an immediately and long-lasting breakdown of the trajectory in Slavonski Brod-Posavina region. By way of example, our synthetic control estimates show that per capita GDP in Zagreb decreased by around 40 percent by the end of Croatia's war of independence in 1995. A rapidly ensuing recovery is vividly indicated by the convergence of the observed per capita GDP to the level implied by Zagreb's synthetic control group. By contrast, the estimated average per capita GDP decline in Slavonski Brod-Posavina is 41 percent with no sign of temporary deviation. By the end of our sample, Herzegovina Neretva's per capita GDP is 50 percent below the level of its synthetic peer without armed conflict in the pre- and post-intervention period, suggesting that the negative effect does not seem to fade away and only increases in its magnitude up to the present day.

The synthetic control groups for Croatia are primarily dominated by attributes of Japanese prefectures, Chines provinces, German and Brazilian federal states as well as Venezuelan and Mexican states among several others. For example, Zagreb pre-war per capita GDP per capita trajectory is best reproduced as a convex combination of the implied characteristics of Schleswig Holstein (39%), Tabasco (17%), Osaka (15%), Brandenburg (9%), Amazonas (7%), Tokyo (5%), Colima (4%), Champagne-Ardenne (3%), Estado Falcon (1%) as well as Baja California-Sur and Centre Val de Loire with a very minor weight in the full composition. On the other hand, the per capita GDP trajectory of Slavonski Brod-Posavina before the outbreak of the armed conflict is best synthesized by a weighted average of Okinawa (22%), Tibet (19%), Ceara (16%), Sichuan and Chongqing (8%), Yucatan (8%), Thüringen (7%), Amazonas (5%), Tabasco (4%), Estado de Mexíco (4%), Yamanashi (3%), Schleswig Holstein (2%), Chiba (1%), Delta Amacuro (<1%), Niedersachsen (<1%) and Estado Fálcon (<1%), respectively. Against the backdrop of the effect heterogeneity, our evidence shows that Zagreb City experienced a more temporary economic disruption whereas non-capital regions suffered a sharp breakdown in pre-conflict growth trajectories, with a much slower recovery. The extent of economic loss was amplified in areas heavily exposed to armed insurgencies, suggesting an arguably strong association between conflict intensity and long-term economic stagnation.

### S.1.3 North Macedonia

The economic consequences of armed conflict in the former Yugoslavia reveal profound and persistent disruptions, with North Macedonia emerging as a striking case of economic collapse despite comparatively lower casualties. Our analysis estimates that, on average, war-induced economic contractions resulted in a per capita GDP approximately 61% lower than the counterfactual scenario absent of conflict. Unlike other former Yugoslav republics, both capital and non-capital regions of North Macedonia experienced a deep and sustained breakdown in economic growth. Notably, Skopje recovered its absolute pre-war GDP level before the 2008 financial crisis but consistently lagged behind its synthetic control group, with no signs of convergence. Its post-war trajectory mirrors the economic attributes of diverse global regions, including Okinawa (20%), Paraiba (16%), and Yucatán (10%). The Northeast Macedonia-Kumanovo region stands out as the most severely affected, suffering an estimated 73% decline in per capita GDP relative to its synthetic counterpart. This region has yet to reclaim its pre-war economic level, with actual per capita GDP stagnating at approximately $5,000, a stark contrast to the $18,000 projected by the synthetic control method. By way of example, Kumanovo's pre-war economic growth trajectory is best reproduced as a convex combination of the growth and development attributes of Okinawa (46%), Sichuan with Chonqing (45%), Estado Fálcon (5%), Estado de Mexico (5%), and Baja California Sur (<1%), respectively. These findings underscore the long-term economic cost inflicted by armed conflict, illustrating how war can not only disrupt immediate economic activity but also fundamentally derail regional growth trajectories over decades. Figure 14 presents the estimated per capita GDP effect of armed conflict in former Yugoslavia from regional growth trajectories of Skopje and Northeast Macedonia-Kumanovo.

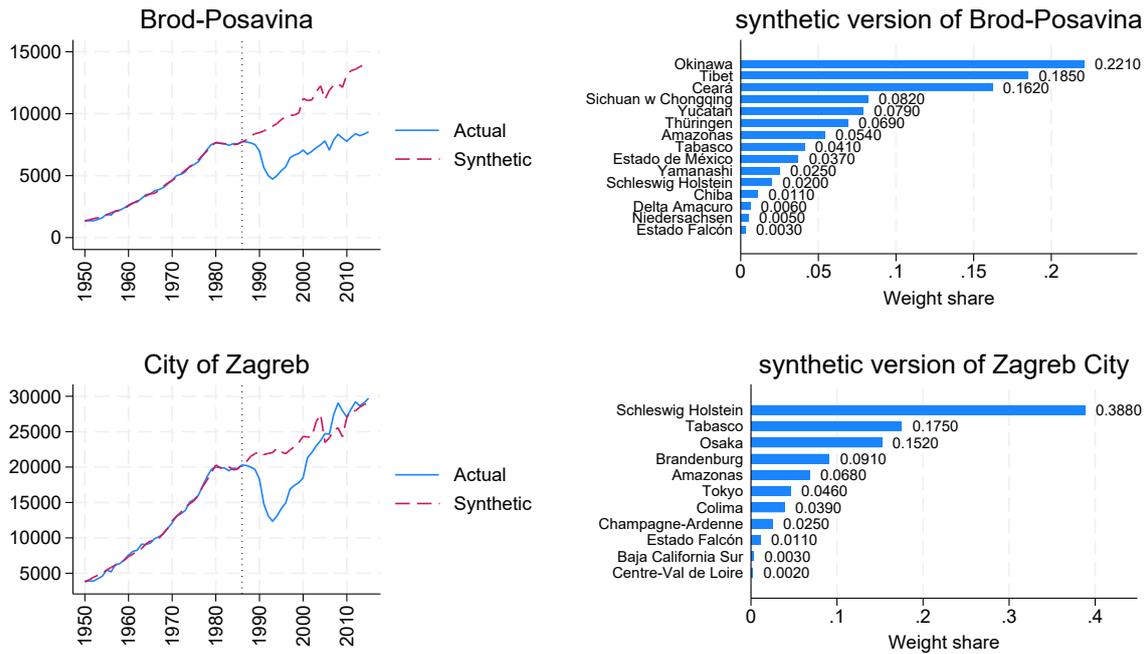

**Figure 13:** *Long-term effect of Yugoslav war on economic growth trajectories of Croatian counties, 1950-2015*

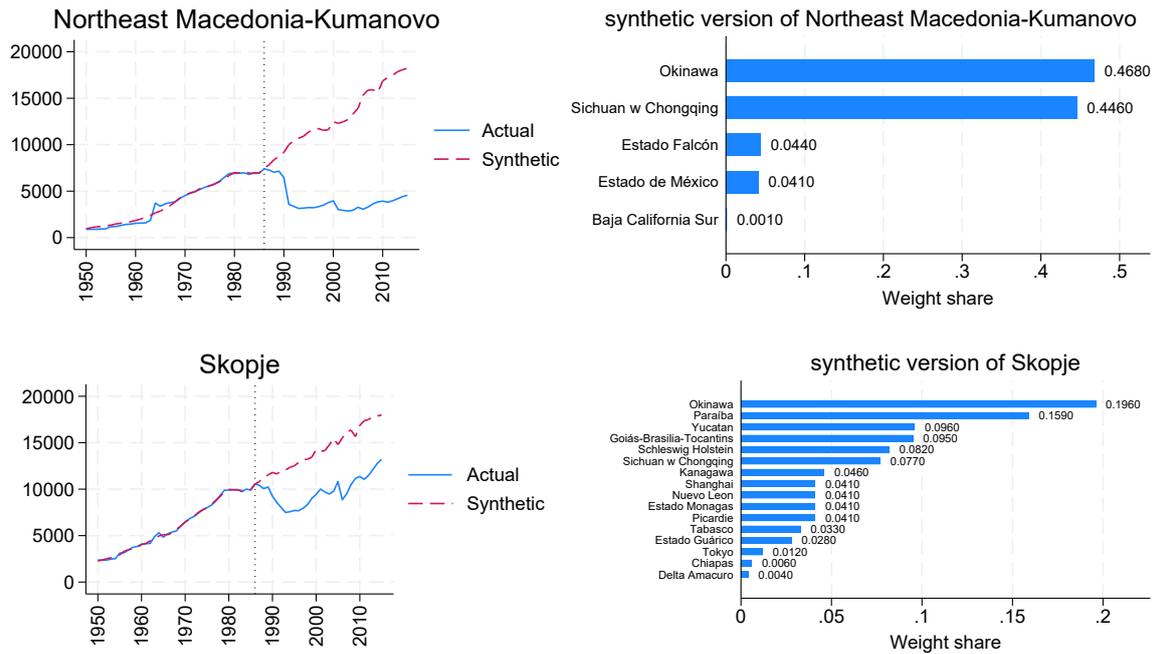

**Figure 14:** *Long-term effect of Yugoslav war on economic growth trajectories of North Macedonian regions, 1950-2015*

### S.1.4 Serbia

The economic consequences of armed conflict in Serbia reveal a profound and enduring downturn, exacerbated by the post-Dayton Agreement period, NATO's 1999 bombing campaign, and prolonged domestic political instability. Our analysis estimates that, on average, Serbia's per capita GDP remains 66% lower than its synthetic peer in the hypothetical absence of war, with both the capital and non-capital regions suffering a deep and persistent breakdown in economic growth—unlike the post-war trajectories observed in Bosnia, Croatia, and Slovenia. The case of Belgrade illustrates the compounding effects of war and external military intervention. Initially on a path of economic recovery following the Dayton Agreement, the capital's trajectory appears to have been derailed by NATO airstrikes, preventing it from regaining its absolute pre-war GDP level. Instead, Belgrade fell increasingly behind its synthetic control group, with an estimated 74% lower per capita GDP in the aftermath of the war and a persistent 62% gap at the end of the sample period.

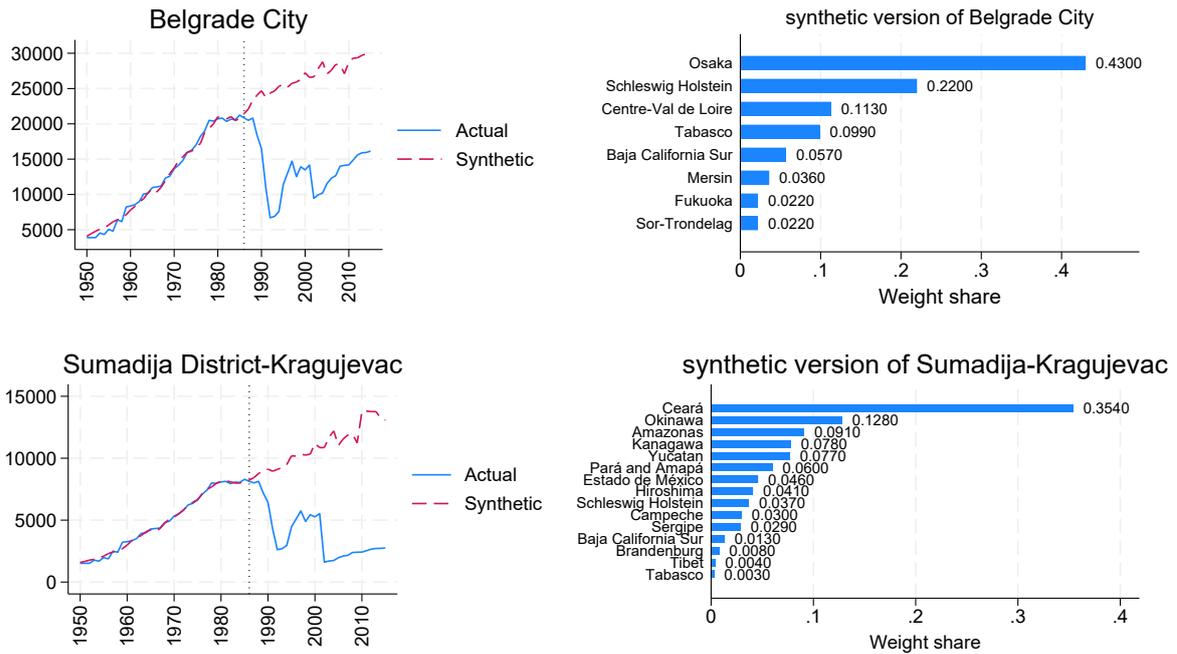

**Figure 15:** *Long-term effect of Yugoslav war on economic growth trajectories of Serbian counties, 1950-2015*

Its economic trajectory aligns most closely with a synthetic composition dominated by Osaka (43%), Schleswig-Holstein (22%), and Centre-Val de Loire (11%), among others. The most severely affected region, Šumadija-Kragujevac, experienced unparalleled economic devastation, with per capita GDP stagnating at $2,700, a dramatic 79% decline from its estimated synthetic counterpart of $13,000. The region's pre-war growth pattern is best approximated by a weighted combination of Ceará (35%), Okinawa (13%), and Amazonas (10%), reflecting an economic trajectory that diverged sharply from its synthetic counterpart. These findings underscore not only the long-term economic scarring inflicted by armed conflict but also the compounding effects of international intervention and political instability, which have left Serbia on a markedly divergent growth path relative to its hypothetical non-war scenario. Figure 15 presents the estimated per capita GDP effect of armed conflict in former Yugoslavia from regional growth trajectories of Belgrade and Šumadija District.

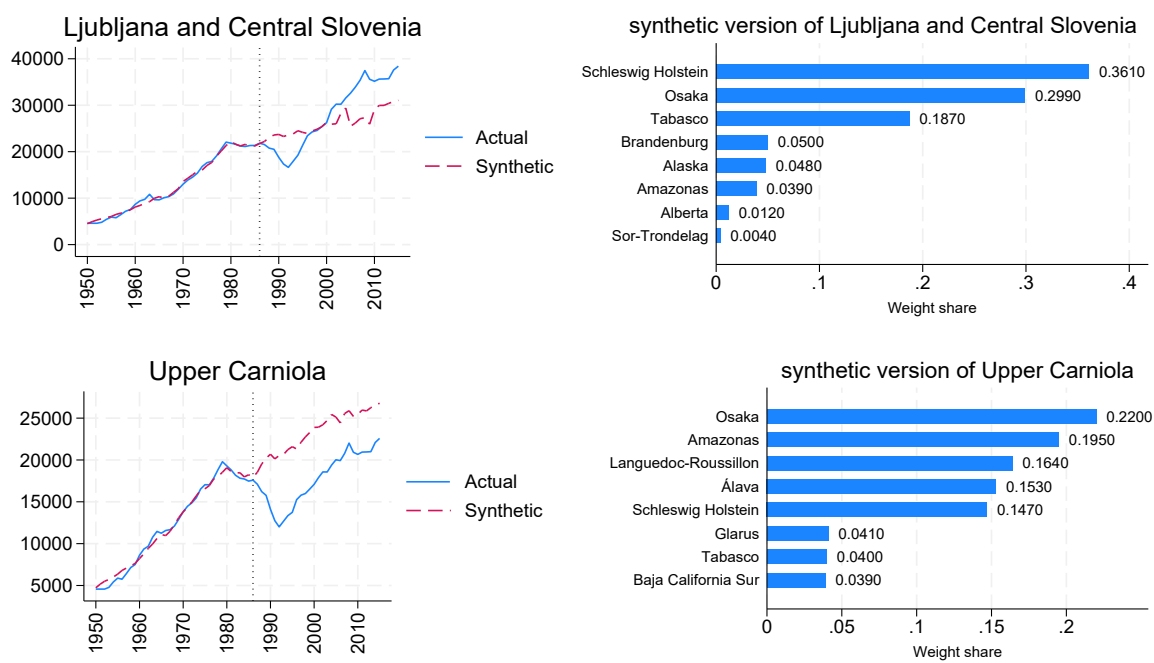

**Figure 16:** *Long-term effect of Yugoslav war on economic growth trajectories of Serbian counties, 1950-2015*

### S.1.5   Slovenia

Among the former Yugoslav republics, Slovenia stands out as the case where the economic effects of war were most transitory, with the notable exception of three regions (Central Sava, Mura, Upper Carniola). Unlike in other successor states, the war did not induce a permanent economic breakdown but rather a temporary negative deviation from its counterfactual growth trajectory. On average, region-level per capita GDP declined by approximately 3% relative to its synthetic control group. However, by the early 2000s, the vast majority of regions fully closed the per capita GDP gap and subsequently outperformed their synthetic peers. This suggests that, in contrast to the prolonged economic decline observed elsewhere in former Yugoslavia, the war and disintegration catalysed structural and institutional reforms in Slovenia, accelerating economic transformation towards a modern European economy. Ljubljana and Central Slovenia exemplify this kind of catalysed trajectory, with their per capita GDP surpassing the synthetic control by 21% by the end of the sample period.

Ljubljana's growth trajectory is best reproduced by a weighted combination of Schleswig-Holstein (31%), Osaka (30%), Tabasco (19%), and other regions with minor weights. Our interpretation is that Slovenia's rapid accession to the European Union and OECD, coupled with its macroeconomic stability, played a pivotal role in closing the GDP per capita gap and ultimately driving long-run gains beyond the counterfactual scenario.

However, not all Slovenian regions experienced a swift and sustained recovery. Upper Carniola, one of the most industrialized and economically advanced regions of pre-war Yugoslavia, diverged from this broader trend, undergoing a long-lasting disruption in economic growth. Our estimates indicate that, on average, Upper Carniola's per capita GDP remains 29% lower than its synthetic counterpart (p-value = 0.000), with a still significant 17% gap at the end of the sample period. This persistent divergence from its synthetic peer underscores regional heterogeneity in post-war economic trajectories and reflects Upper Carniola's pre-war stronger structural alignment with Western Europe, as well as its historical industrial strength compared to the regions elsewhere in former Yugoslavia.

The region's pre-conflict economic path is most closely approximated by a synthetic combination of Osaka (22%), Amazonas (20%), Languedoc-Roussillon (16%), Álava (15%), and others, indicating a strong resemblance to a diverse set of highly industrialized and export-oriented economies. The broader donor pool for Slovenia's synthetic control groups—comprising predominantly German, French, Japanese, Mexican, Swiss, and Venezuelan regions—reflects the region's unique structural positioning within both Western European and global economic networks. These findings highlight the complex and regionally differentiated impact of war and disintegration, demonstrating how the same historical shock can yield divergent long-term economic consequences depending on pre-existing economic structures, integration pathways, and institutional adaptability. Figure 16 presents the estimated counterfactual scenarios of economic growth for the least and most severely affected Slovenian regions.

## S.2 Matching on salient features of Yugoslav economy using Mediterranean donor pool

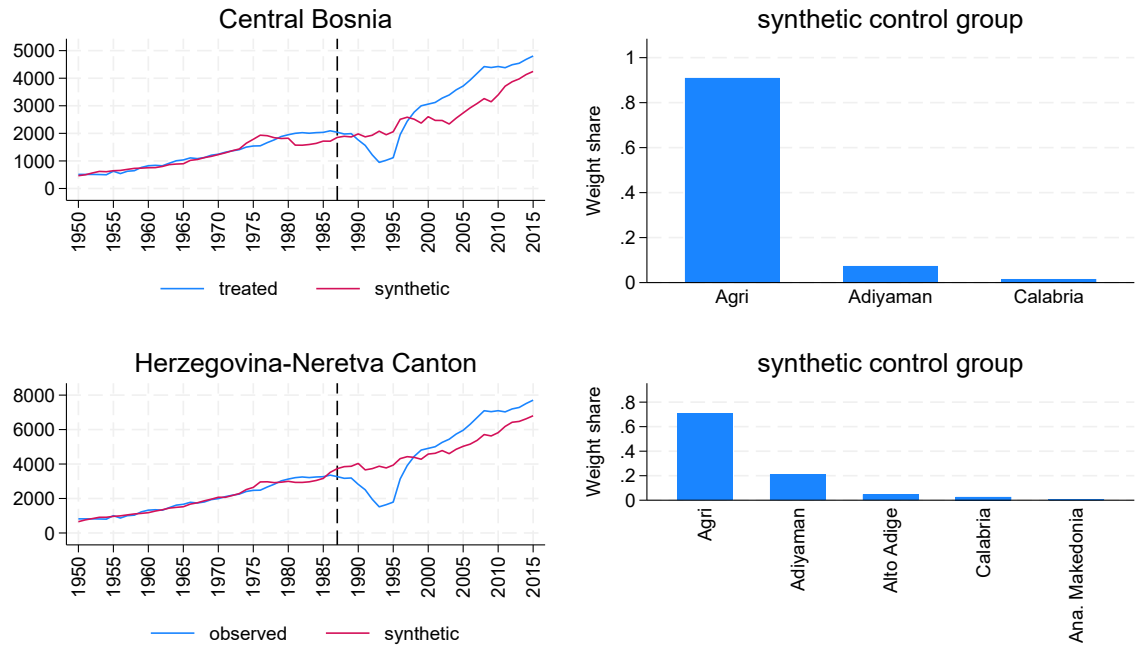

**Figure 17:** *The Long-term effect of Yugoslav war on economic growth trajectories of least and worst impacted Bosnian cantons, 1950-2015*

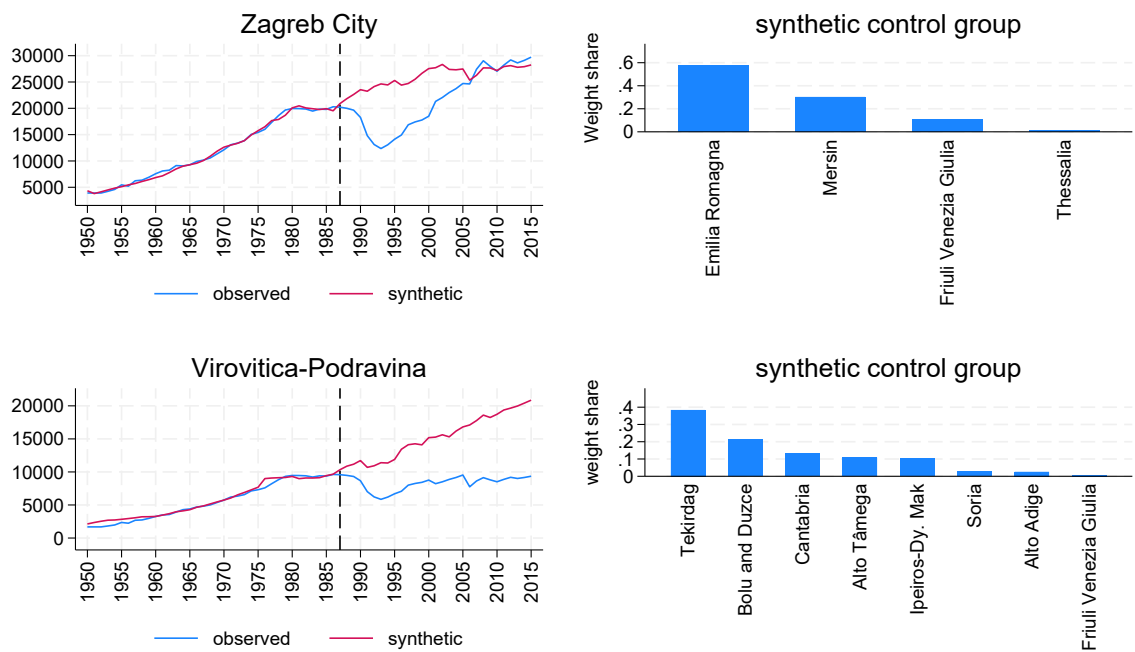

**Figure 18:** *The Long-term effect of Yugoslav war on economic growth trajectories of least and worst impacted Croatian counties, 1950-2015*

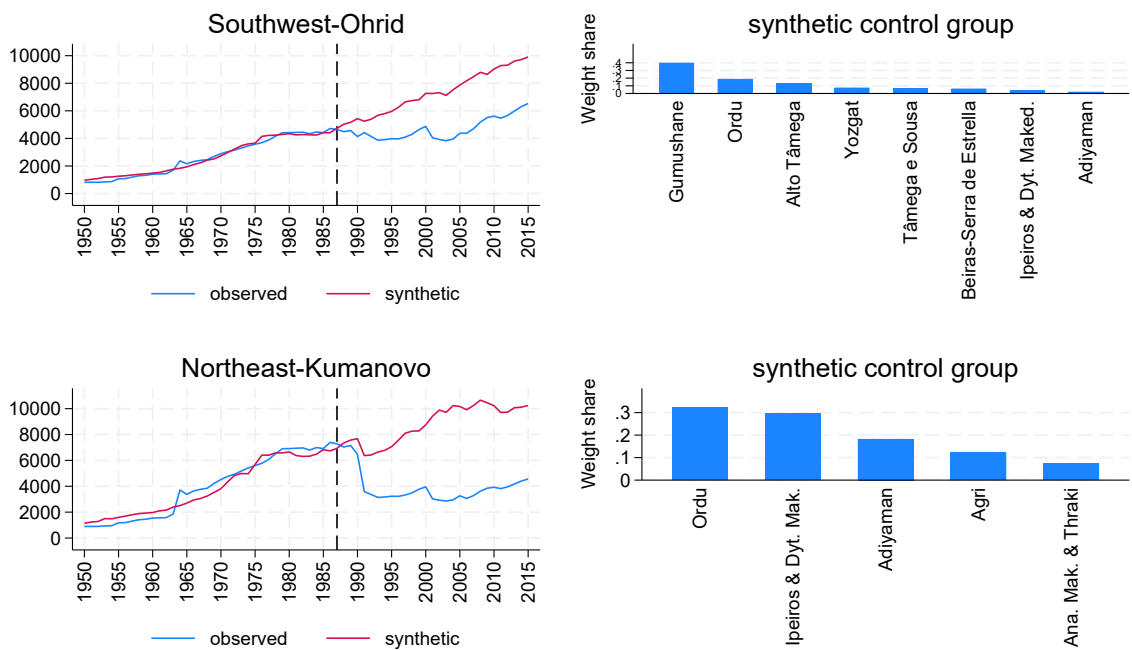

**Figure 19:** *The Long-term effect of Yugoslav war on economic growth trajectories of least and worst impacted North Macedonian regions, 1950-2015*

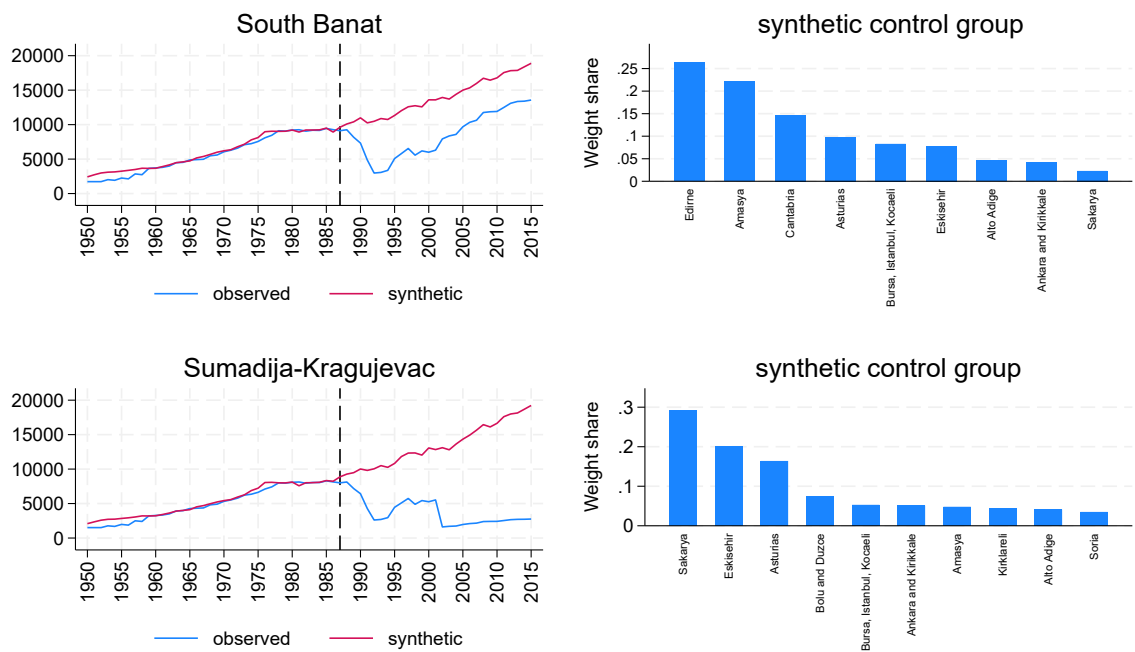

**Figure 20:** *The Long-term effect of Yugoslav war on economic growth trajectories of least and worst impacted Serbian districts, 1950-2015*

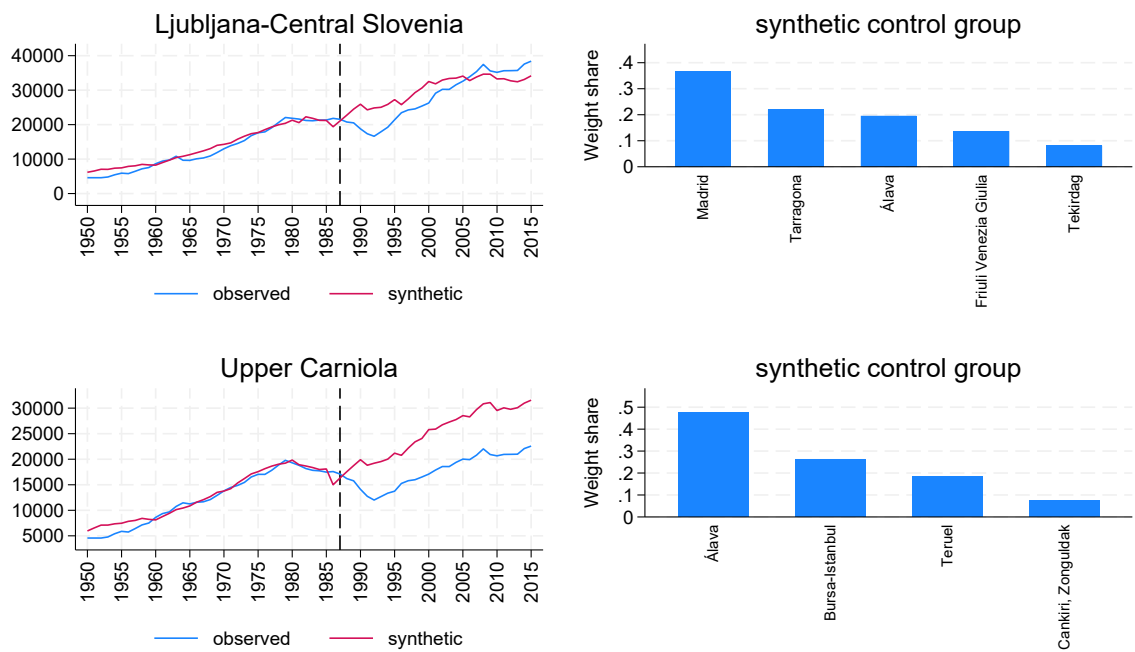

**Figure 21:** *The Long-term effect of Yugoslav war on economic growth trajectories of least and worst impacted Slovenian regions, 1950-2015*